\documentclass[11pt]{article}
\usepackage{amssymb,amsmath,amsfonts}
\usepackage{graphicx}
\usepackage{graphics}
\usepackage{eepic,epsfig}
\usepackage{dcolumn}

\@addtoreset{equation}{section}

\makeatother

\begin{document}

\title{Two-point functions and the vacuum densities \\
in the Casimir effect for the Proca field}
\author{A. A. Saharian\thanks{%
Corresponding author, E-mail: saharian@ysu.am }\, and H. H. Asatryan \\
\textit{Institute of Physics, Yerevan State University, }\\
\textit{1 Alex Manoogian Street, 0025 Yerevan, Armenia }}
\maketitle

\begin{abstract}
We investigate the properties of the vacuum state for the Proca field in the
geometry of two parallel plates on background of (D+1)-dimensional Minkowski
spacetime. The two-point functions for the vector potential and the field
tensor are evaluated for higher-dimensional generalizations of the perfect
magnetic conductor (PMC) and perfect electric conductor (PEC) boundary
conditions. Explicit expressions are provided for the vacuum expectation
values (VEVs) of the electric and magnetic field squares, field condensate,
and for the VEV of the energy-momentum tensor. In the zero-mass limit the
VEVs of the electric and magnetic field squares and the condensate reduce to
the corresponding expressions for a massless vector field. The same is the
case for the VEV of the energy-momentum tensor in the problem with PEC
conditions. However, for PMC conditions the zero-mass limit for the vacuum
energy-momentum tensor differs from the corresponding VEV for a massless
field. This difference in the zero-mass limits is related to the different
influences of the boundary conditions on the longitudinal polarization mode
of a massive vector field. The PMC conditions constrain all the polarization
modes including the longitudinal mode, whereas PEC conditions do not
influence the longitudinal mode. The vacuum energy-momentum tensor is
diagonal. The normal stress is uniformly distributed in the region between
the plates and vanishes in the remaining regions. The corresponding Casimir
forces are attractive for both boundary conditions. The Casimir-Polder
forces acting on a polarizable particle are repulsive with respect to the
nearest plate for PMC conditions. For PEC conditions those forces are
attractive for $D\geq 2$ and vanish for $D=1$. For $D\geq 3$, the vacuum
energy density is positive for the PMC conditions and negative for the PEC
conditions. For $D=2$ and for PMC conditions, the vacuum energy density is
negative in the region between the plates and vanishes in the remaining
regions. For $D=1$ and PMC conditions, the energy density is negative
everywhere. In the case of PEC conditions, the vacuum energy density is
positive for $D=2$ and vanishes for $D=1$.
\end{abstract}

\bigskip

Keywords: Casimir effect, Proca field, perfect conductor boundary conditions

\bigskip

\section{Introduction}

In several problems in quantum field theory, the interaction of a field with
external stimuli can be described by an effective model, in which boundary
or periodicity conditions are imposed on the field operator. These
conditions affect the spectrum of quantum fluctuations of the field,
resulting in changes to the expectation values of the physical observables
that characterize the given state. The dependence of physical quantities on
specified conditions is known by the general name Casimir effect (for
reviews see \cite{Most97,Milt02,Bord09,Casi11}). This effect has been
investigated for various fields, boundary conditions, boundary and
background geometries. The great interest in the Casimir effect is due to
its important role in condensed matter physics, quantum field theory,
gravitation, and cosmology.

In the literature, the Casimir effect has been studied for fields with
different spins. This has been done for both massless and massive fields. In
the case of massive fields, the expectation values of physical quantities
associated with the Casimir effect typically decay exponentially when the
distance of the observation point from the boundary exceeds the Compton
wavelength. However, in background gravitational fields, the behavior at
large distances can exhibit qualitative differences. Such behavior is
present, for example, in the Casimir effect on the background of de Sitter
spacetime \cite{Saha09,Eliz10,Milt12}. The decay of the vacuum densities for
both planar and spherical boundaries, as functions of the distance from the
boundary, is the power law for both massless and massive fields. Depending
on the value of the field mass, the fall-off can be monotonic or damping
oscillatory.

Motivated by applications in electromagnetism, the Casimir effect for vector
fields has been studied mainly for the massless case. However, massive
vector fields appear in a number of physical models. Of particular relevance
are the gauge vector fields within the Standard Model. In electrodynamic
systems, a photon can also acquire an effective mass, and the corresponding
dynamics can be described in terms of a massive vector field model. These
include electrodynamic models in media and theories with extra dimensions.
The motivation to investigate quantum phenomena in the context of massive
vector fields is also attributed to the potential non-zero mass of the
photon (for constraints on the photon mass see, e.g., \cite{Tu05,Gold10}).
Two approaches have been discussed in the literature to introduce the mass
for vector fields. In the first one, the term $m^{2}A_{\mu }A^{\mu }$ is
added to the Maxwellian Lagrangian for the electromagnetic field with the
vector potential $A_{\mu }$. This term breaks the local gauge invariance of
the theory. The corresponding field is known in the literature as the Proca
vector field (sometimes also as the de Broglie-Proca field) \cite{Proc36}.
In the second approach, suggested in \cite{Stue38,Stue38b} (for a review see 
\cite{Rueg04}), in addition to the mass term an additional scalar field $%
\varphi $ is introduced in the Lagrangian density, directly interacting with
the vector field through the term $-mA^{\mu }\partial _{\mu }\varphi $. The
gauge transformation of the scalar field $\varphi $ compensates the gauge
breaking part coming from the mass term $m^{2}A_{\mu }A^{\mu }$ and the
local gauge invariance is preserved. The corresponding theory is referred to
as Stueckelberg massive electromagnetism. Note that the Stueckelberg
Lagrangian is obtained from the Lagrangian for the Proca field by the
replacement $A_{\mu }\rightarrow A_{\mu }+2\sqrt{\pi }\partial _{\mu
}\varphi /m$.

The nonzero mass of the vector field will give two types of modifications in
the expressions for physical characteristics in the Casimir effect. The
first one comes from the additional longitudinal polarization state and the
second one is the modification in the contribution from transverse
polarizations. Two possibilities in the standard Casimir setup for two
parallel plates have been discussed in \cite{Davi81}. They correspond to
complete reflection and perfect transmission for the modes of longitudinal
polarization. More general problems with finite-width dielectric plates have
been discussed in \cite{Bart85}. The Proca field modes and corresponding
boundary conditions on the separation boundaries are presented in detail.
The generalization for magnetodielectric plates at zero and finite
temperatures is discussed in \cite{Teo10,Teo12}. The Casimir energy for a
massive vector field in the geometry of perfectly conducting concentric
spherical bodies is studied in \cite{Teo11}. The Casimir-Polder interactions
with massive photons have been investigated in \cite{Matt19}. In models of
extra compact spatial dimensions, the effective dynamics of 4D vector field
is described in terms of the Proca Lagrangian with the mass determined by
the length of compact dimensions. The Casimir effect in a simple 5D model
with the extra dimension compactified to a circle is considered in \cite%
{Eder08}. The model with extra dimension of the Randall-Sundrum type has
been discussed in \cite{Teo10b}. The Casimir effect for Stueckelberg massive
vector field is studied in \cite{Belo16}. Note that the precise measurements
of the Casimir forces provide sensitive tests of new physics beyond the
Standard Model (see, e.g., \cite{Bord09,Decc07,Klim17,Klim21} and references
therein)

Previous research on the Casimir effect for the Proca field has focused on
global characteristics of the vacuum in 3D space, such as the total vacuum
energy and the forces acting on the boundaries with electric type
conditions. Models with an extra dimension ($D=4$) have also been considered
incorporating 3D and 4D boundary conditions on planar boundaries. In the
present paper, the local characteristics of the vacuum state for the
geometry of parallel plates in a general number of spatial dimensions $D$
are investigated. Two types of boundary conditions are discussed, which
generalize the $D=3$ perfect magnetic conductor (PMC) and perfect electric
conductor (PEC) boundary conditions to the general case of spatial
dimension. The expectation values of local physical observables, biliniear
in the field, are obtained from the two-point functions and their
derivatives in the coincidence limit of the spacetime arguments. These
functions will be evaluated for both types of boundary conditions.

The paper is organized as follows. In the next section, we describe the
problem setup and present the mode functions for a massive Proca field
obeying the PMC conditions on two parallel plates in a general number of
spatial dimensions. The two-point Wightman functions for the vector
potential and field tensor are considered in Section \ref{sec:TPvector}. By
using these functions, the VEVs of local physical observables are
investigated in Section \ref{sec:VEVs}. Aiming to compare the VEVs in the
zero-mass limit of the Proca field with the corresponding VEVs for a
massless vector field, the two-point functions and the expectation values
for the latter are discussed in Section \ref{sec:Massless}. The Casimir
densities in the geometry of parallel plates for the Proca field with PEC
conditions are investigated in Section \ref{sec:BCcond}. The main results of
the paper are summarized in section \ref{sec:Conc}. The details of the
intermediate evaluation of the two-point functions are described in Appendix %
\ref{sec:App1}.

\section{Proca field modes for PMC boundary conditions}

\label{sec:Modes}

We consider massive Proca vector field $A_{\mu }(x)$ in $(D+1)$-dimensional
flat spacetime with the metric tensor $g_{\mu \nu }=\mathrm{diag}%
(1,-1,\ldots ,-1)$ and the coordinates $(x^{0}=t,x^{1},\ldots
,x^{D-1},x^{D}=z)$. The corresponding action is given by%
\begin{equation}
S=\int d^{D+1}x\,L(A_{\mu }),\;L(A_{\mu })=-\frac{F_{\mu \nu }F^{\mu \nu
}-2m^{2}A_{\mu }A^{\mu }}{16\pi },  \label{S}
\end{equation}%
where $F_{\mu \nu }=\partial _{\mu }A_{\nu }-\partial _{\nu }A_{\mu }$ is
the field tensor. The variation of the action with respect to the vector
field leads to the field equation 
\begin{equation}
\nabla _{\nu }F^{\mu \nu }-m^{2}A^{\mu }=0.  \label{Feq}
\end{equation}%
By taking into account that for an antisymmetric tensor $B^{\mu \nu }$ one
has $\nabla _{\mu }\nabla _{\nu }B^{\mu \nu }=0$, from (\ref{Feq}) it
follows that $m^{2}\nabla _{\mu }A^{\mu }=0$. Therefore, for a massive field
($m\neq 0$) we have 
\begin{equation}
\partial _{\mu }A^{\mu }=0.  \label{CondA}
\end{equation}%
For a massless field this is a gauge condition on the vector potential that
is independent from the field equation. Substituting in (\ref{Feq}) the
expression for the field tensor one obtains the field equation in terms of
the vector potential: 
\begin{equation}
\left( \nabla _{\nu }\nabla ^{\nu }+m^{2}\right) A^{\mu }=0.  \label{Feq3}
\end{equation}

We are interested in the influence of two parallel plates, located at $z=0$
and $z=a$, on the local properties of the vacuum state for the field $A_{\mu
}(x)$. In this section, we assume that on the plates the field obeys the
boundary condition 
\begin{equation}
n^{\nu }F_{\mu \nu }=0,\;z=0,a,  \label{BC}
\end{equation}%
where $n^{\nu }=\delta _{D}^{\nu }$ is the normal to the plates. The
boundary condition (\ref{BC}) is a higher dimensional generalization of PMC
boundary conditions in Maxwell's electromagnetism (for the realization of
PMC boundary conditions by using metamaterials see, for example, \cite%
{Lope22}). In terms of the $D=3$ electric and magnetic fields $\mathbf{E}$
and $\mathbf{B}$, they are written in the form $\mathbf{n}\cdot \mathbf{E}=0$
and $\mathbf{n}\times \mathbf{B}=0$. The condition (\ref{BC}) is also used
in bag models for hadrons for confining the vector fields in finite regions.
The properties of the vacuum state are encoded in two-point functions and
for evaluation of those functions we will employ a direct summation over the
complete set of vector modes obeying the boundary condition (\ref{BC}). The
plates at $z=0$ and $z=a$ divide the space into three spatial regions: $z<0$%
, $0<z<a$, and $z>a$. The following discussion will mainly concern the
region between the plates. The two-point functions and the expectation
values in other regions are obtained by the limiting transition $%
a\rightarrow \infty $ while keeping the position of one of the plates fixed.

By taking into account the problem symmetry, we present the modes for the
vector field in the form%
\begin{equation}
A_{\mu }=f_{\mu }(z)e^{-ik_{l}x_{\parallel }^{l}},  \label{Amu}
\end{equation}%
with the components of the wave vector $k_{l}$ and $x_{\parallel
}^{l}=(x^{0},x^{1},\ldots ,x^{D-1})$. Here and below, the latin indices take
the values $l=0,1,\ldots ,D-1$. From the field equation (\ref{Feq}) with $%
\mu =D$ it follows that 
\begin{equation}
ik_{l}\partial _{D}f^{l}+k_{D}^{2}f^{D}=0,  \label{EqD}
\end{equation}%
where 
\begin{equation}
k_{D}^{2}=k_{l}k^{l}-m^{2}=\omega ^{2}-\mathbf{k}^{2}-m^{2},\;\omega
=k^{0},\;\mathbf{k}=(k^{1},k^{2},\ldots ,k^{D-1}).  \label{lam}
\end{equation}%
The equation (\ref{CondA}) gives the relation 
\begin{equation}
k_{l}f^{l}=-i\partial _{z}f^{D}.  \label{CondAf}
\end{equation}%
Plugging this in (\ref{EqD}) one obtains $\partial
_{z}^{2}f^{D}+k_{D}^{2}f^{D}=0$ with the general solution%
\begin{equation}
f^{D}(z)=c_{1}^{D}e^{ik^{D}z}+c_{2}^{D}e^{-ik^{D}z}.  \label{fD}
\end{equation}%
The field equation (\ref{Feq}) with $\mu =l$, by taking into account (\ref%
{CondAf}), leads to the equation $\partial _{z}^{2}f^{l}+k_{D}^{2}f^{l}=0$,
having the general solution 
\begin{equation}
f^{l}(z)=c_{1}^{l}e^{ik^{D}z}+c_{2}^{l}e^{-ik^{D}z},\;l=0,1,\ldots ,D-1.
\label{fl}
\end{equation}%
The coefficients in (\ref{fD}) and (\ref{fl}) are connected by the relation (%
\ref{CondAf}):%
\begin{equation}
k_{l}c_{1}^{l}=k^{D}c_{1}^{D},\;k_{l}c_{2}^{l}=-k^{D}c_{2}^{D}.  \label{Relc}
\end{equation}

For the field tensor corresponding to the solutions of the form (\ref{Amu})
one has%
\begin{align}
F^{nl} &= i\left( k^{l}f^{n}-k^{n}f^{l}\right) e^{-ik_{l}x_{\parallel }^{l}},
\notag \\
F^{Dl} &= \left( -\partial _{z}f^{l}+i\,k^{l}f^{D}\right)
e^{-ik_{l}x_{\parallel }^{l}}.  \label{FDl}
\end{align}%
The boundary condition (\ref{BC}) is reduced to the relations%
\begin{equation}
\left( \,k^{l}c_{1}^{D}-k^{D}c_{1}^{l}\right) e^{ik^{D}z}+\left(
\,k^{l}c_{2}^{D}+k^{D}c_{2}^{l}\right) e^{-ik^{D}z}=0,\;z=0,a.  \label{BCcl1}
\end{equation}%
There are two classes of modes. The first one corresponds to transverse
modes with $c_{1}^{D}=c_{2}^{D}=0$ and $k_{l}c_{1}^{l}=k_{l}c_{2}^{l}=0$.
For them from (\ref{BCcl1}) we find $c_{1}^{l}=c_{2}^{l}$. The corresponding
eigenvalues for $k^{D}$ are the roots of the equation $\sin (k^{D}a)=0$ and
they are given by $k^{D}=\pi n/a$ with $n=1,2,\ldots $. The second class of
modes corresponds to the longitudinal mode with $f^{D}\neq 0$. For it,
multiplying (\ref{BCcl1}) by $k_{l}$, summing over $l$, and by taking into
account the relations (\ref{Relc}), we get%
\begin{equation*}
m^{2}\left( c_{1}^{D}e^{ik^{D}z}+c_{2}^{D}e^{-ik^{D}z}\right) =0,\;z=0,a.
\end{equation*}%
From here we see that $c_{2}^{D}=-c_{1}^{D}$ and, again, the eigenvalues are
roots of the equation $\sin (k^{D}a)=0$ with $k^{D}=\pi n/a$, $n=1,2,\ldots $%
. In addition, there is also a zero mode with $k^{D}=0$. For this mode, as
it follows from (\ref{BCcl1}), $f^{D}(z)=0$.

The complete set of the vector modes is specified by the set of quantum
numbers $\beta =(\mathbf{k},k^{D}=\pi n/a,\sigma )$, where $\sigma
=1,2,\ldots ,D$ enumerates the polarization degrees of freedom. Introducing
the polarization vector $\epsilon _{(\sigma )}^{l}$, from the analysis given
above it follows that for the transverse modes with polarizations $\sigma
=1,2,\ldots ,D-1$ the corresponding vector potential is given by 
\begin{equation}
A_{(\beta )l}=C_{\beta }\epsilon _{(\sigma )l}\cos \left( k^{D}z\right)
e^{-ik_{j}x_{\parallel }^{j}},\;A_{(\beta )D}=0.  \label{Atr}
\end{equation}%
For the polarization vector one has $k^{l}\epsilon _{(\sigma )l}=0$ and 
\begin{equation}
g^{il}\epsilon _{(\sigma ^{\prime })i}\epsilon _{(\sigma )l}=-\delta
_{\sigma ^{\prime }\sigma }.  \label{eps1}
\end{equation}%
It obeys the relation%
\begin{equation}
\sum_{\sigma =1}^{D-1}\epsilon _{(\sigma )i}\epsilon _{(\sigma )l}=\frac{%
k_{i}k_{l}}{\lambda ^{2}}-g_{il},\;\lambda ^{2}=k_{l}k^{l}=k_{D}^{2}+m^{2}.
\label{eps2}
\end{equation}%
In (\ref{Atr}), the eigenvalues of $k^{D}$ are given by 
\begin{equation}
k^{D}=\frac{\pi n}{a},\;n=0,1,2,\ldots ,  \label{kD}
\end{equation}%
and they include also the zero mode with $n=0$.

The longitudinal mode corresponds to the polarization $\sigma =D$ and for
the corresponding vector potential one has%
\begin{align}
A_{(\beta )l} &= -\frac{k_{l}k^{D}}{\lambda ^{2}}C_{\beta }\cos \left(
k^{D}z\right) e^{-ik_{j}x_{\parallel }^{j}},  \notag \\
A_{(\beta )D} &= iC_{\beta }\sin \left( k^{D}z\right) e^{-ik_{j}x_{\parallel
}^{j}},  \label{Al}
\end{align}%
where the eigenvalues for $k^{D}$ are given by (\ref{kD}) with $n=1,2,\ldots 
$. Note that for this mode $F_{(\beta )ls}=\partial _{l}A_{(\beta
)s}-\partial _{s}A_{(\beta )l}=0$.

The vector modes are normalized by the condition 
\begin{equation}
\int d^{D}x\,\left[ A_{(\beta ^{\prime })\mu }^{\ast }\partial _{0}A_{(\beta
)}^{\mu }-\left( \partial _{0}A_{(\beta ^{\prime })\mu }^{\ast }\right)
A_{(\beta )}^{\mu }\right] =4i\pi \delta _{\beta \beta ^{\prime }},
\label{NC1}
\end{equation}%
which, for the modes (\ref{Atr}) and (\ref{Al}) is reduced to 
\begin{equation}
\int d^{D}x\,A_{(\beta ^{\prime })\mu }^{\ast }A_{(\beta )}^{\mu }=-\frac{%
2\pi }{\omega }\delta _{\beta \beta ^{\prime }}.  \label{NC2}
\end{equation}%
For the polarizations $\sigma =1,2,\ldots ,D-1$ this gives 
\begin{equation}
\left\vert C_{\beta }\right\vert ^{2}=\frac{2\left( 1-\delta _{0n}/2\right) 
}{\left( 2\pi \right) ^{D-2}a\omega },  \label{Cbet1}
\end{equation}%
and for the longitudinal polarization ($\sigma =D$):%
\begin{equation}
\left\vert C_{\beta }\right\vert ^{2}=\frac{2\lambda ^{2}}{\left( 2\pi
\right) ^{D-2}m^{2}a\omega }.  \label{Cbet2}
\end{equation}%
Note that for the latter the modes are singular in the zero mass limit.

\section{Two-point functions for PMC conditions}

\label{sec:TPvector}

In this section we consider the two-point functions for the vector potential
and the field tensor.

\subsection{Two-point functions for the vector potential}

The two-point function for the vector potential (the positive-frequency
Wightman function) is defined as the expectation value%
\begin{equation}
\left\langle 0\right\vert A_{\mu }(x)A_{\nu }(x^{\prime })\left\vert
0\right\rangle \equiv \left\langle A_{\mu }A_{\nu }^{\prime }\right\rangle ,
\label{AA1}
\end{equation}%
where $\left\vert 0\right\rangle $ is the vacuum state. Having the complete
set of modes $A_{(\beta )\mu }(x)$, it is evaluated by using the mode-sum
formula%
\begin{equation}
\left\langle A_{\mu }A_{\nu }^{\prime }\right\rangle =\int d\mathbf{k}%
\,\sum_{\sigma =1}^{D}\sum_{n=0}^{\infty }A_{(\beta )\mu }(x)A_{(\beta )\nu
}^{\ast }(x^{\prime }).  \label{AA2}
\end{equation}%
The summation over the polarizations $\sigma =1,2,\ldots ,D-1$ is done by
using the formula (\ref{eps2}) and the components of the two-point functions
are presented in the form%
\begin{align}
\left\langle A_{\mu }A_{\nu }^{\prime }\right\rangle & =\left( \frac{%
\partial _{\mu }\partial _{\nu }^{\prime }}{m^{2}}-g_{\mu \nu }\right)
A(x,x^{\prime }),  \notag \\
\left\langle A_{D}A_{D}^{\prime }\right\rangle & =\frac{\partial
_{D}\partial _{D}^{\prime }}{m^{2}}A(x,x^{\prime })+B(x,x^{\prime }),
\label{ADD}
\end{align}%
where $\mu +\nu <2D$. Here we have introduced the notations%
\begin{align}
A(x,x^{\prime })& =\frac{2}{a}\int \frac{d\mathbf{k}}{\left( 2\pi \right)
^{D-2}}\,\sideset{}{'}{\sum}_{n=0}^{\infty }\frac{1}{\omega }\cos \left(
k^{D}z\right) \cos \left( k^{D}z^{\prime }\right) e^{-ik_{l}\Delta
x_{\parallel }^{l}},  \notag \\
B(x,x^{\prime })& =\frac{2}{a}\int \frac{d\mathbf{k}}{\left( 2\pi \right)
^{D-2}}\,\sum_{n=1}^{\infty }\frac{1}{\omega }\sin \left( k^{D}z\right) \sin
\left( k^{D}z^{\prime }\right) e^{-ik_{l}\Delta x_{\parallel }^{l}},
\label{Bxx}
\end{align}%
where $\Delta x_{\parallel }^{l} =x_{\parallel }^{l} -x_{\parallel }^{\prime
l} $ and the prime on the summation sign means that the term $n=0$ should be
taken with additional coefficient 1/2. For the further transformation it is
convenient to present the two-point functions (\ref{Bxx}) in the form%
\begin{equation}
A(x,x^{\prime })=\sum_{j=\mp 1}A_{j}(x,x^{\prime }),\;B(x,x^{\prime
})=-\sum_{j=\mp 1}jA_{j}(x,x^{\prime }),  \label{Abxx}
\end{equation}%
with the function%
\begin{equation}
A_{j}(x,x^{\prime })=\frac{1}{a}\int \frac{d\mathbf{k}}{\left( 2\pi \right)
^{D-2}}\,\sum_{n=0}^{\infty \prime }\frac{1}{\omega }\cos \left[
k^{D}(z+jz^{\prime })\right] e^{-ik_{l}\Delta x_{\parallel }^{l}}.
\label{Ajxx}
\end{equation}%
The representation (\ref{Ajxx4}) for this function, more convenient in the
evaluation of the VEVs, is given in Appendix \ref{sec:App1}. By using that
representation, for the functions (\ref{Bxx}) we get 
\begin{align}
A(x,x^{\prime })& =\frac{2m^{D-1}}{\left( 2\pi \right) ^{\frac{D-1}{2}}}%
\sum_{j=\mp 1}\sum_{n=-\infty }^{+\infty }f_{\frac{D-1}{2}%
}(mb_{j,n}(x,x^{\prime })),  \notag \\
B(x,x^{\prime })& =-\frac{2m^{D-1}}{\left( 2\pi \right) ^{\frac{D-1}{2}}}%
\sum_{j=\mp 1}\sum_{n=-\infty }^{+\infty }jf_{\frac{D-1}{2}%
}(mb_{j,n}(x,x^{\prime })),  \label{Bxx2}
\end{align}%
where the notation%
\begin{equation}
b_{j,n}(x,x^{\prime })=\sqrt{\left( 2na-z-jz^{\prime }\right) ^{2}-\Delta
x_{l}\Delta x^{l}},  \label{bjn}
\end{equation}%
is introduced. In (\ref{Bxx2}), we have defined the function 
\begin{equation}
f_{\nu }(x)=x^{-\nu }K_{\nu }(x),  \label{fnu}
\end{equation}%
with $K_{\nu }(x)$ being the modified Bessel function of the second kind.

Substituting (\ref{Bxx2}) into (\ref{ADD}) we find%
\begin{equation}
\left\langle A_{\mu }A_{\nu }^{\prime }\right\rangle =\frac{2m^{D-1}}{\left(
2\pi \right) ^{\frac{D-1}{2}}}\sum_{j=\mp 1}\sum_{n=-\infty }^{+\infty
}A_{\mu \nu }^{(j,n)}(x,x^{\prime }),  \label{AA3}
\end{equation}%
where%
\begin{align}
A_{il}^{(j,n)}(x,x^{\prime })& =g_{il}Df_{\frac{D+1}{2}}(mb_{j,n}(x,x^{%
\prime }))-m^{2}\left( g_{il}b_{j,n}^{2}(x,x^{\prime })+\Delta x_{i}\Delta
x_{l}\right) f_{\frac{D+3}{2}}(mb_{j,n}(x,x^{\prime })),  \notag \\
A_{Dl}^{(j,n)}(x,x^{\prime })& =-m^{2}\left( 2na-z-jz^{\prime }\right)
\Delta x_{l}f_{\frac{D+3}{2}}(mb_{j,n}(x,x^{\prime })),  \notag \\
A_{DD}^{(j,n)}(x,x^{\prime })& =j\left[ Df_{\frac{D+1}{2}}(b_{j,n}(x,x^{%
\prime }))+m^{2}\Delta x_{l}\Delta x^{l}f_{\frac{D+3}{2}}(mb_{j,n}(x,x^{%
\prime }))\right] ,  \label{ADD2}
\end{align}%
and $A_{lD}^{(j,n)}(x,x^{\prime })=-jA_{Dl}^{(j,n)}(x,x^{\prime })$. In
deriving these representations we have used the relation $f_{\nu }^{\prime
}(x)=-xf_{\nu +1}(x)$. Another relation for the function $f_{\nu }(x)$
directly follows from the recurrence relation for the modified Bessel
function:%
\begin{equation}
x^{2}f_{\nu +1}(x)=2\nu f_{\nu }(x)+f_{\nu -1}(x).  \label{relfnu}
\end{equation}%
The two-point function (\ref{AA3}) is decomposed into three contributions:%
\begin{equation}
\left\langle A_{\mu }A_{\nu }^{\prime }\right\rangle =\left\langle A_{\mu
}A_{\nu }^{\prime }\right\rangle _{0}+\left\langle A_{\mu }A_{\nu }^{\prime
}\right\rangle _{1}+\left\langle A_{\mu }A_{\nu }^{\prime }\right\rangle
_{2},  \label{AAdec}
\end{equation}%
where%
\begin{align}
\left\langle A_{\mu }A_{\nu }^{\prime }\right\rangle _{0}& =\frac{2m^{D-1}}{%
\left( 2\pi \right) ^{\frac{D-1}{2}}}A_{\mu \nu }^{(-1,0)}(x,x^{\prime }), 
\notag \\
\left\langle A_{\mu }A_{\nu }^{\prime }\right\rangle _{1}& =\frac{2m^{D-1}}{%
\left( 2\pi \right) ^{\frac{D-1}{2}}}A_{\mu \nu }^{(+1,0)}(x,x^{\prime }),
\label{AApart} \\
\left\langle A_{\mu }A_{\nu }^{\prime }\right\rangle _{2}& =\frac{2m^{D-1}}{%
\left( 2\pi \right) ^{\frac{D-1}{2}}}\sum_{j=\mp 1}\sum_{n=-\infty ,\neq
0}^{+\infty }A_{\mu \nu }^{(j,n)}(x,x^{\prime }).  \notag
\end{align}%
The part $\left\langle A_{\mu }A_{\nu }^{\prime }\right\rangle _{0}$
corresponds to the two-point function in the boundary-free geometry with $%
-\infty <z,z^{\prime }<+\infty $. The sum of the two first terms in the
right-hand side of (\ref{AAdec}) presents the two-point function in the
region $z,z^{\prime }>0$ for the problem with a single plate at $z=0$.
Hence, the part $\left\langle A_{\mu }A_{\nu }^{\prime }\right\rangle _{1}$
is interpreted as the contribution to the two-point function induced by the
presence of a single plate at $z=0$. Finally, the term $\left\langle A_{\mu
}A_{\nu }^{\prime }\right\rangle _{2}$ is induced by the second plate at $%
z=a $ when we add it to the geometry of a single plate at $z=0$. Note that
the part in (\ref{AA3}) with $A_{\mu \nu }^{(+1,1)}(x,x^{\prime })$
corresponds to the contribution of a single plate at $z=a$ (the plate at $%
z=0 $ is absent) in the region $z,z^{\prime }<a$.

As it is seen from (\ref{AA3}), the two-point functions of the vector
potential are singular in the zero mass limit. In the expressions for the
VEVs of physical observables the two-point functions enter in the form of
the product $m^{2}\left\langle A_{\mu }A_{\nu }^{\prime }\right\rangle $.
The zero mass limit of this product is finite and is given by%
\begin{equation}
\left[ m^{2}\left\langle A_{\mu }A_{\nu }^{\prime }\right\rangle \right]
_{m\rightarrow 0}=\frac{2\Gamma \left( \frac{D+1}{2}\right) }{\pi ^{\frac{D-1%
}{2}}}\sum_{j=\mp 1}\sum_{n=-\infty }^{+\infty }\frac{A_{(0)\mu \nu
}^{(j,n)}(x,x^{\prime })}{b_{j,n}^{D+1}(x,x^{\prime })},  \label{AA3m0}
\end{equation}%
with the notations%
\begin{align}
A_{(0)il}^{(j,n)}(x,x^{\prime })& =-g_{il}-\left( D+1\right) \frac{\Delta
x_{i}\Delta x_{l}}{b_{j,n}^{2}(x,x^{\prime })},  \notag \\
A_{(0)Dl}^{(j,n)}(x,x^{\prime })& =-\left( D+1\right) \Delta x_{l}\frac{%
2na-z-jz^{\prime }}{b_{j,n}^{2}(x,x^{\prime })},  \notag \\
A_{(0)DD}^{(j,n)}(x,x^{\prime })& =jD+j\left( D+1\right) \frac{\Delta
x_{l}\Delta x^{l}}{b_{j,n}^{2}(x,x^{\prime })}.  \label{ADDm0}
\end{align}%
The consequences for the nonzero zero-mass limit for physical observables
bilinear in the field operator will be discussed below.

\subsection{Two-point function for the field tensor}

The VEVs of the electric and magnetic field squares are obtained from the
two-point functions for the field tensor in the coincidence limit of the
spacetime arguments. These two-point functions are defined as%
\begin{equation}
\left\langle 0\right\vert F_{\mu \nu }(x)F_{\rho \sigma }(x^{\prime
})\left\vert 0\right\rangle =\left\langle F_{\mu \nu }F_{\rho \sigma
}^{\prime }\right\rangle ,  \label{FFxx}
\end{equation}%
with the expansion over the modes%
\begin{equation}
\left\langle F_{\mu \nu }F_{\rho \sigma }^{\prime }\right\rangle =\int d%
\mathbf{k}\,\sum_{\sigma =1}^{D}\sum_{n=0}^{\infty }F_{(\beta )\mu \nu
}(x)F_{(\beta )\rho \sigma }^{\ast }(x^{\prime }).  \label{FFxx1}
\end{equation}%
The mode functions have the form 
\begin{align}
F_{(\beta )lp}& =-iC_{\beta }\left( k_{l}\epsilon _{(\sigma
)p}-k_{p}\epsilon _{(\sigma )l}\right) \cos \left( k^{D}z\right)
e^{-ik_{l}x_{\parallel }^{l}},  \notag \\
F_{(\beta )Dl}& =-C_{\beta }k^{D}\epsilon _{(\sigma )l}\sin \left(
k^{D}z\right) e^{-ik_{l}x_{\parallel }^{l}},  \label{FDlm}
\end{align}%
for the polarizations $\sigma =1,2,\ldots ,D-1$ and 
\begin{equation}
F_{(\beta )il}=0,\;F_{(\beta )Dl}=-C_{\beta }m^{2}\frac{k_{l}}{\lambda ^{2}}%
\sin \left( k^{D}z\right) e^{-ik_{j}x_{\parallel }^{j}},  \label{FDlm2}
\end{equation}%
for the polarization $\sigma =D$. The normalization constants are given by (%
\ref{Cbet1}) and (\ref{Cbet2}). Note that, unlike the mode functions for the
vector potential, the mode functions for the field tensor are regular in the
zero-mass limit. For the longitudinal polarization ($\sigma =D$) the
electric field is directed along the axis $x^{D}$ and it vanishes on the
plates.

With the mode functions (\ref{FDlm}) and (\ref{FDlm2}), from (\ref{FFxx1})
the following expressions are obtained%
\begin{align}
\left\langle F_{ik}F_{lp}^{\prime }\right\rangle & =2\left[ g_{p[i}\partial
_{k]}\partial _{l}^{\prime }+g_{l[k}\partial _{i]}\partial _{p}^{\prime }%
\right] A(x,x^{\prime }),  \notag \\
\left\langle F_{Di}F_{kl}^{\prime }\right\rangle & =2g_{i[k}\partial
_{l]}^{\prime }\partial _{D}A(x,x^{\prime }),  \notag \\
\left\langle F_{Di}F_{Dl}^{\prime }\right\rangle & =\partial _{i}\partial
_{l}^{\prime }B(x,x^{\prime })-g_{il}\partial _{D}\partial _{D}^{\prime
}A(x,x^{\prime }),  \label{FDiFDl}
\end{align}%
in terms of the functions (\ref{Bxx2}). Here the square brackets in the
expressions of indices mean the antisymmetrization over the enclosed indices:%
\begin{equation}
a_{[i}b_{k]}=\frac{1}{2}\left( a_{i}b_{k}-a_{k}b_{i}\right) .
\label{antisym}
\end{equation}%
Note that the longitudinal modes contribute only to the two-point function $%
\left\langle F_{Di}F_{Dl}^{\prime }\right\rangle $.

By using the representation (\ref{Bxx2}) we get%
\begin{equation}
\left\langle F_{\mu \nu }F_{\rho \sigma }^{\prime }\right\rangle =-\frac{%
4m^{D+1}}{\left( 2\pi \right) ^{\frac{D-1}{2}}}\sum_{j=\mp 1}\sum_{n=-\infty
}^{\infty }F_{\mu \nu \rho \sigma }^{(j,n)}(x,x^{\prime }),  \label{FF2p1}
\end{equation}%
with the functions%
\begin{align}
F_{iklp}^{(j,n)}(x,x^{\prime }) &= 2g_{p[i}g_{k]l}f_{\frac{D+1}{2}%
}(mb_{j,n}(x,x^{\prime }))  \notag \\
&+m^{2}\left[ g_{p[i}\Delta x_{k]}\Delta x_{l}-g_{l[i}\Delta x_{k]}\Delta
x_{p}\right] f_{\frac{D+3}{2}}(mb_{j,n}(x,x^{\prime })),  \notag \\
F_{Dikl}^{(j,n)}(x,x^{\prime }) &= m^{2}g_{i[k}\Delta x_{l]}\left(
2na-z-jz^{\prime }\right) f_{\frac{D+3}{2}}(mb_{j,n}(x,x^{\prime })),
\label{FF2p} \\
F_{DiDl}^{(j,n)}(x,x^{\prime }) &= j\left\{ -g_{il}f_{\frac{D+1}{2}%
}(mb_{j,n}(x,x^{\prime }))\right.  \notag \\
& +\left. \frac{m^{2}}{2}\left[ g_{il}\left( 2na-z-jz^{\prime }\right)
^{2}-\Delta x_{i}\Delta x_{l}\right] f_{\frac{D+3}{2}}(mb_{j,n}(x,x^{\prime
}))\right\} .  \notag
\end{align}%
Similar to the two-point function for the vector potential, the two-point
function (\ref{FF2p1}) is decomposed into the contributions corresponding to
the boundary-free geometry (the part with $F_{\mu \nu \rho \sigma
}^{(-1,0)}(x,x^{\prime })$), to the part induced by a single plate at $z=0$
in the region $z,z^{\prime }>0$ (the part with $F_{\mu \nu \rho \sigma
}^{(+1,0)}(x,x^{\prime })$), and to the part induced by the plate at $z=a$
in the region $0<z,z^{\prime }<a$ when we add it to the geometry with a
single plate at $z=0$ (the remaining part in (\ref{FF2p1})).

From (\ref{FF2p1}) it follows that the two-point function for the field
tensor is finite in the zero-mass limit. It is given by the formula 
\begin{equation}
\left\langle F_{\mu \nu }F_{\rho \sigma }^{\prime }\right\rangle
_{m\rightarrow 0}=-4\frac{\Gamma \left( \frac{D+1}{2}\right) }{\pi ^{\frac{%
D-1}{2}}}\sum_{j=\mp 1}\sum_{n=-\infty }^{\infty }\frac{F_{(0)\mu \nu \rho
\sigma }^{(j,n)}(x,x^{\prime })}{b_{j,n}^{D+1}(x,x^{\prime })},
\label{FFm0lim}
\end{equation}%
where the functions%
\begin{align}
F_{(0)iklp}^{(j,n)}(x,x^{\prime })& =2g_{p[i}g_{k]l}+\left( D+1\right) \frac{%
g_{p[i}\Delta x_{k]}\Delta x_{l}-g_{l[i}\Delta x_{k]}\Delta x_{p}}{%
b_{j,n}^{2}(x,x^{\prime })},  \notag \\
F_{(0)Dikl}^{(j,n)}(x,x^{\prime })& =\left( D+1\right) g_{i[k}\Delta x_{l]}%
\frac{2na-z-jz^{\prime }}{b_{j,n}^{D+3}(x,x^{\prime })},  \label{FF2pm0} \\
F_{(0)DiDl}^{(j,n)}(x,x^{\prime })& =-jg_{il}+j\frac{D+1}{2}\frac{%
g_{il}\left( 2na-z-jz^{\prime }\right) ^{2}-\Delta x_{i}\Delta x_{l}}{%
b_{j,n}^{2}(x,x^{\prime })},  \notag
\end{align}%
are introduced for the separate components.

\section{VEVs of the field squares and energy-momentum tensor}

\label{sec:VEVs}

\subsection{General expressions}

The VEVs of the observables bilinear in the field operator are obtained from
the two-point functions given above in the coincidence limit $x^{\prime
}\rightarrow x$. This limit is divergent. For points away from boundaries
the divergences come from the part in the two-point functions corresponding
to the boundary-free geometry. The important advantage of the
representations given above is that the boundary-free contribution (denoted
by $\left\langle \cdots \right\rangle _{0}$) is explicitly separated. The
renormalized VEVs are obtained by subtracting from the two-point functions
the parts presenting the boundary-free geometry and then taking the
coincidence limit $x^{\prime }\rightarrow x$. All of the VEVs discussed
below are symmetric with respect to the hyperplane $z=a/2$. This feature is
a direct consequence of the problem symmetry.

We start the consideration from the VEV of the electric field squared. The
corresponding renormalized VEV is given by%
\begin{equation}
\left\langle E^{2}\right\rangle =-g^{\mu \nu }\lim_{x^{\prime }\rightarrow
x} \left[ \left\langle F_{0\mu }F_{0\nu }^{\prime }\right\rangle
-\left\langle F_{0\mu }F_{0\nu }^{\prime }\right\rangle _{0}\right] .
\label{E2}
\end{equation}%
By taking into account the expressions (\ref{FF2p1}), (\ref{FF2p}) and using
the relation (\ref{relfnu}) for the two-point functions in the region $0<z<a$%
, we get%
\begin{align}
\left\langle E^{2}\right\rangle & =-\frac{2m^{D+1}}{\left( 2\pi \right) ^{%
\frac{D-1}{2}}}\left\{ 2\sum_{n=1}^{\infty }\left[ \left( D-1\right) f_{%
\frac{D+1}{2}}(2nma)-f_{\frac{D-1}{2}}(2nma)\right] \right.  \notag \\
& \left. +\sum_{n=-\infty }^{\infty }\left[ 3\left( D-1\right) f_{\frac{D+1}{%
2}}(2mz_{n})+f_{\frac{D-1}{2}}(2mz_{n})\right] \right\} ,  \label{E2r}
\end{align}%
with the notation%
\begin{equation}
z_{n}=|z-na|.  \label{zn}
\end{equation}%
The $n=0$ term in (\ref{E2r}) corresponds to the VEV of the electric field
squared in the geometry with a single plate $z=0$:%
\begin{equation}
\left\langle E^{2}\right\rangle _{1}=-\frac{2m^{D+1}}{\left( 2\pi \right) ^{%
\frac{D-1}{2}}}\left[ 3\left( D-1\right) f_{\frac{D+1}{2}}(2m|z|)+f_{\frac{%
D-1}{2}}(2m|z|)\right] .  \label{E21}
\end{equation}%
This expression is valid in both regions $z>0$ and $z<0$. The VEV (\ref{E21}%
) is negative. The expression for the electric field squared in the region $%
z>a$ is obtained from (\ref{E21}) by the replacement $z\rightarrow z-a$. By
taking into account that the function $f_{\nu }(x)$ is monotonically
decreasing with increasing $x$, it can be shown that%
\begin{equation}
\sum_{n=-\infty }^{\infty }f_{\nu }(2mz_{n})>2\sum_{n=1}^{\infty }f_{\nu
}(2ma).  \label{Ineq}
\end{equation}%
From here it follows that the VEV (\ref{E2r}) is negative as well. Hence,
the VEV of the electric field squared for the PMC conditions is negative
everywhere.

Another important bilinear combination of the field is the invariant 
\begin{equation}
\left\langle F_{\mu \nu }F^{\mu \nu }\right\rangle =g^{\mu \rho }g^{\nu
\sigma }\lim_{x^{\prime }\rightarrow x}\left[ \left\langle F_{\mu \nu
}F_{\rho \sigma }^{\prime }\right\rangle -\left\langle F_{\mu \nu }F_{\rho
\sigma }^{\prime }\right\rangle _{0}\right] .  \label{FFr}
\end{equation}%
This VEV is the analog of the gluon condensate in quantum chromodynamics.
Combining the formulas for the separate components of the two-point
functions for the field tensor, one finds%
\begin{align}
\left\langle F_{\mu \nu }F^{\mu \nu }\right\rangle =&\frac{4Dm^{D+1}}{\left(
2\pi \right) ^{\frac{D-1}{2}}}\left\{ -2\sum_{n=1}^{\infty }f_{\frac{D-1}{2}%
}(2nma)\right.  \notag \\
&+\left. \sum_{n=-\infty }^{\infty }\left[ 2(D-1)f_{\frac{D+1}{2}%
}(2mz_{n})+f_{\frac{D-1}{2}}(2mz_{n})\right] \right\} .  \label{FFr2}
\end{align}%
This VEV is positive. For the VEV of the Lagrangian density we get%
\begin{equation}
\left\langle L(A_{\mu })\right\rangle =-\frac{Dm^{D+1}}{\left( 2\pi \right)
^{\frac{D+1}{2}}}\sum_{n=-\infty }^{\infty }\left[ Df_{\frac{D+1}{2}%
}(2mz_{n})+f_{\frac{D-1}{2}}(2mz_{n})\right] .  \label{Lag}
\end{equation}%
The VEVs $\left\langle F_{\mu \nu }F^{\mu \nu }\right\rangle $ and $%
\left\langle L(A_{\mu })\right\rangle $ in the region $z<0$ coincide with
the corresponding terms with $n=0$ in (\ref{FFr2}) and (\ref{Lag}). The
expressions in the region $z>a$ are obtained from those in $z<0$ by the
replacement $z\rightarrow z-a$.

The VEV of the magnetic field squared, $\left\langle B^{2}\right\rangle $,
is obtained from the relation 
\begin{equation}
\left\langle B^{2}\right\rangle =\left\langle E^{2}\right\rangle +\frac{1}{2}%
\left\langle F_{\mu \nu }F^{\mu \nu }\right\rangle .  \label{B20}
\end{equation}%
This leads to the formula%
\begin{align}
\left\langle B^{2}\right\rangle = & -\frac{2\left( D-1\right) m^{D+1}}{%
\left( 2\pi \right) ^{\frac{D-1}{2}}}\left\{ 2\sum_{n=1}^{\infty }\left[ f_{%
\frac{D+1}{2}}(2nma)+f_{\frac{D-1}{2}}(2nma)\right] \right.  \notag \\
& \left. -\sum_{n=-\infty }^{\infty }\left[ (2D-3)f_{\frac{D+1}{2}%
}(2mz_{n})+f_{\frac{D-1}{2}}(2mz_{n})\right] \right\} .  \label{B2}
\end{align}%
In the geometry of a single plate at $z=0$ we have%
\begin{equation}
\left\langle B^{2}\right\rangle _{1}=\frac{2\left( D-1\right) m^{D+1}}{%
\left( 2\pi \right) ^{\frac{D-1}{2}}}\left[ (2D-3)f_{\frac{D+1}{2}%
}(2m|z|)+f_{\frac{D-1}{2}}(2m|z|)\right] .  \label{B21}
\end{equation}%
Note that the VEV of the magnetic field squared is zero for $D=1$. The VEV
in the region $z>a$ is obtained from (\ref{B21}) making the replacement $%
z\rightarrow z-a$. For $D\geq 2$ the VEV of the magnetic field squared in
the regions $z<0$ and $z>a$ is negative for PMC boundary conditions on the
plate. The same is the case for the region $0<z<a$. That can be easily seen
from (\ref{B2}) by using the relation (\ref{Ineq}).

Now we turn to the VEV of the energy-momentum tensor. For the Proca field
the corresponding operator is given by the expression%
\begin{equation}
T_{\mu \nu }=-\frac{g^{\alpha \rho }F_{(\mu \alpha }F_{\nu )\rho
}-m^{2}A_{(\mu }A_{\nu )}}{4\pi }-g_{\mu \nu }L(A_{\mu }),  \label{Tmu}
\end{equation}%
where the Lagrangian density is given by (\ref{S}) and the brackets in the
expression for indices mean the symmetrization over the enclosed indices. By
using the expressions for the VEVs of the separate parts in (\ref{Tmu}),
given above, we see that the VEVs of the off-diagonal components vanish. For
the diagonal components we get (no summation over $l$)%
\begin{align}
\left\langle T_{l}^{l}\right\rangle & =\frac{m^{D+1}}{\left( 2\pi \right) ^{%
\frac{D+1}{2}}}\left\{ -2D\sum_{n=1}^{\infty }f_{\frac{D+1}{2}}(2nma)+\left(
D-2\right) \right.  \notag \\
& \times \left. \sum_{n=-\infty }^{\infty }\left[ \left( D-1\right) f_{\frac{%
D+1}{2}}(2mz_{n})+f_{\frac{D-1}{2}}(2mz_{n})\right] \right\} ,  \notag \\
\left\langle T_{D}^{D}\right\rangle & =\frac{2Dm^{D+1}}{\left( 2\pi \right)
^{\frac{D+1}{2}}}\sum_{n=1}^{\infty }\left[ Df_{\frac{D+1}{2}}(2nma)+f_{%
\frac{D-1}{2}}(2nma)\right] ,  \label{TDD}
\end{align}%
where $l=0,1,\ldots ,D-1$. This relations show that the vacuum energy
density ($l=0$) is equal to the stresses along the directions parallel to
the plates. Of course, this property is a consequence of the invariance of
the problem with respect to the Lorentz boosts along the directions parallel
to the plates. The vacuum energy density is negative for $D\leq 2$ and
positive for $D\geq 3$. Note that for $D=2$ the distribution of the vacuum
energy density in the region between the plates is uniform. The normal
stress $\left\langle T_{D}^{D}\right\rangle $ does not depend on the $z$%
-coordinate and it vanishes in the geometry of a single plate. This property
could be directly obtained from the conservation equation $\partial _{\nu
}\left\langle T^{\mu \nu }\right\rangle =0$.

The components of the vacuum energy-momentum tensor in the geometry of a
single plate at $z=0$ are given by the $n=0$ term in (\ref{TDD}) (no
summation over $l$):%
\begin{equation}
\left\langle T_{l}^{l}\right\rangle _{1}=\frac{\left( D-2\right) m^{D+1}}{%
\left( 2\pi \right) ^{\frac{D+1}{2}}}\left[ \left( D-1\right) f_{\frac{D+1}{2%
}}(2m|z|)+f_{\frac{D-1}{2}}(2m|z|)\right] ,  \label{Tll1}
\end{equation}%
for $l=0,1,\ldots ,D-1$, and $\left\langle T_{D}^{D}\right\rangle _{1}=0$.
The corresponding energy density is negative for $D=1$ and positive for $%
D\geq 3$. For $D=2$ the vacuum energy-momentum tensor for PMC boundary
conditions vanishes in the problem with a single plate. The parallel plates
divide the space into three regions. In the regions $z<0$ and $z>a$ the
normal stress vanishes and the energy density and parallel stresses are
given by (\ref{Tll1}) for $z<0$. The expressions for the energy density and
parallel stresses for the region $z>a$ are obtained from (\ref{Tll1}) by the
replacement $z\rightarrow z-a$. In the region between the plates, $0<z<a$,
the VEVs are expressed by the formulas (\ref{TDD}). In this region, the
energy density is negative for $D=1,2$ and positive for $D\geq 3$. The
Casimir force per unit surface of the plates (the Casimir pressure) is
determined by the normal stress as $p_{\mathrm{C}}=-\left\langle
T_{D}^{D}\right\rangle $. It is zero in the regions $z<0$ and $z>a$ and is
negative for the region between the plates. This corresponds to the
attractive Casimir force. Hence, similar to the standard electromagnetic
Casimir effect for parallel conducting plates, the attractive Casimir force
is a consequence of a negative vacuum pressure in the region $0<z<a$.

\subsection{Zero-mass limit of the VEVs}

Let us consider the limit $m\rightarrow 0$ for the VEVs given above. By
taking into account that 
\begin{equation}
f_{\nu }(y)\approx \frac{2^{\nu -1}}{y^{2\nu }}\Gamma (\nu ),
\label{fnusmall}
\end{equation}%
for $0<y\ll 1$, the VEVs of the electric and magnetic field squares are
presented in the form%
\begin{align}
\left\langle E^{2}\right\rangle _{m\rightarrow 0}& =\frac{\left( 1-D\right)
\Gamma \left( \frac{D+1}{2}\right) }{\left( 4\pi \right) ^{\frac{D-1}{2}%
}a^{D+1}}\left[ \zeta (D+1)+\sum_{n=-\infty }^{+\infty }\frac{3/2}{%
\left\vert n-z/a\right\vert ^{D+1}}\right] ,  \notag \\
\left\langle B^{2}\right\rangle _{m\rightarrow 0}& =\frac{\left( 1-D\right)
\Gamma \left( \frac{D+1}{2}\right) }{\left( 4\pi \right) ^{\frac{D-1}{2}%
}a^{D+1}}\left[ \zeta (D+1)+\sum_{n=-\infty }^{+\infty }\frac{3/2-D}{%
\left\vert n-z/a\right\vert ^{D+1}}\right] ,  \label{B2m0}
\end{align}%
where $\zeta (y)=\sum_{n=1}^{\infty }n^{-y}$ is the Riemann zeta function.
For the condensate we get%
\begin{equation}
\left\langle F_{\mu \nu }F^{\mu \nu }\right\rangle _{m\rightarrow 0}=\frac{%
2D\Gamma \left( \frac{D+1}{2}\right) }{\left( 4\pi \right) ^{\frac{D-1}{2}%
}a^{D+1}}\sum_{n=-\infty }^{+\infty }\frac{D-1}{\left\vert n-z/a\right\vert
^{D+1}}.  \label{FFm0}
\end{equation}

The expressions for the nonzero components of the vacuum energy-momentum
tensor in the zero-mass limit read (no summation over $l$)%
\begin{align}
\left\langle T_{l}^{l}\right\rangle _{m\rightarrow 0}& =\frac{D\Gamma \left( 
\frac{D+1}{2}\right) }{\left( 4\pi \right) ^{\frac{D+1}{2}}a^{D+1}}\left[ 
\frac{D-2}{2D}\sum_{n=-\infty }^{\infty }\frac{D-1}{\left\vert
n-z/a\right\vert ^{D+1}}-\zeta (D+1)\right] ,  \notag \\
\left\langle T_{D}^{D}\right\rangle _{m\rightarrow 0}& =\frac{D^{2}\Gamma
\left( \frac{D+1}{2}\right) }{\left( 4\pi \right) ^{\frac{D+1}{2}}a^{D+1}}%
\zeta (D+1).  \label{TDDml0}
\end{align}%
For the corresponding trace one gets 
\begin{equation}
\left\langle T_{l}^{l}\right\rangle _{m\rightarrow 0}=\frac{D\Gamma \left( 
\frac{D+1}{2}\right) }{2\left( 4\pi \right) ^{\frac{D+1}{2}}a^{D+1}}%
\sum_{n=-\infty }^{\infty }\frac{\left( D-1\right) \left( D-2\right) }{%
\left\vert n-z/a\right\vert ^{D+1}}.  \label{Tracem0}
\end{equation}%
The VEV of the energy-momentum tensor in the zero mass limit is traceless in
spatial dimensions $D=1,2$. In spatial dimensions $D\geq 3$ the trace is
nonzero. In the massless limit the action (\ref{S}) is conformally invariant
for $D=3$ and the appearance of the nonzero trace (\ref{Tracem0}) can be
considered as a kind of trace anomaly. It is different from the standard
trace anomaly induced by the curvature of the spacetime (see, e.g., \cite%
{Birr82}).

\begin{figure}[tbph]
\begin{centering}
\epsfig{figure=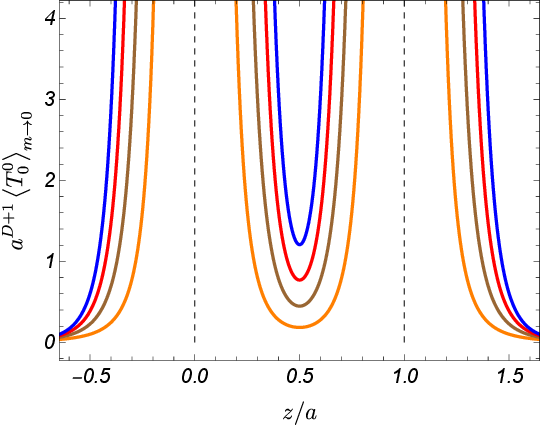,width=7.5cm,height=6.cm}
\par\end{centering}
\caption{Vacuum energy density (in units $1/a^{D+1}$) as a function of $z/a$
for the Proca field in the limit $m\rightarrow 0$. The graphs are plotted
for $D=3,4,5,6$. The product $a^{D+1}\left\langle T_{0}^{0}\right\rangle $
increases with increasing $D$.}
\label{figT00m0}
\end{figure}

Figure \ref{figT00m0} presents the dependence of the vacuum energy density
for the Proca field (measured in units of $1/a^{D+1}$) in the limit $%
m\rightarrow 0$ (see (\ref{TDDml0})) as a function of $z/a$. The graphs are
plotted for $D=3,4,5,6$. The product $a^{D+1}\left\langle
T_{0}^{0}\right\rangle _{m\rightarrow 0}$ increases with increasing $D$. 

\subsection{Asymptotic and numerical analysis}

For points near the plate at $z=0$, $|z|/a\ll 1$, the dominant contribution
to the VEVs $\left\langle E^{2}\right\rangle $, $\left\langle
B^{2}\right\rangle $, and to the vacuum energy density and parallel stresses
comes from the terms $n=0$ in (\ref{E2r}), (\ref{B2}), and (\ref{TDD}). The
leading terms in the asymptotic expansions over $|z|/a$ coincide with the
terms $n=0$ in (\ref{B2m0}) and (\ref{TDDml0}) (no summation over $l$):%
\begin{align}
\left\langle E^{2}\right\rangle & \approx \frac{3\left\langle
B^{2}\right\rangle }{3-2D}\approx -\frac{3\left( D-1\right) \Gamma \left( 
\frac{D+1}{2}\right) }{2\left( 4\pi \right) ^{\frac{D-1}{2}}|z|^{D+1}}, 
\notag \\
\left\langle T_{l}^{l}\right\rangle & \approx \frac{\left( D-2\right) \left(
D-1\right) \Gamma \left( \frac{D+1}{2}\right) }{2\left( 4\pi \right) ^{\frac{%
D+1}{2}}|z|^{D+1}}.  \label{Asnear}
\end{align}%
The leading terms near the plate at $z=a$, assuming that $|z/a-1|\ll 1$, are
obtained from (\ref{Asnear}) making the replacement $|z|\rightarrow |z-a|$.
In particular, the VEV of the electric field squared is negative near the
plates and the energy density is positive for $D\geq 3$.

\begin{figure}[tbph]
\begin{centering}
\epsfig{figure=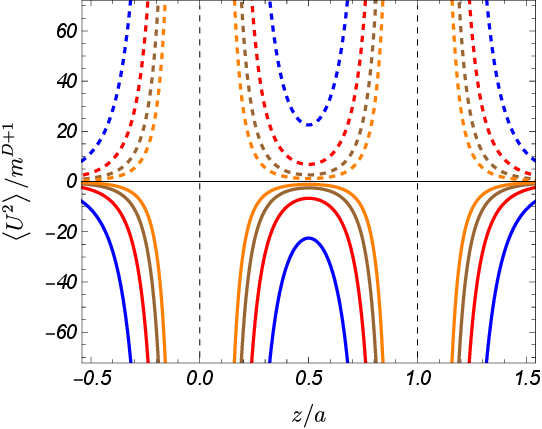,width=7.5cm,height=6.cm}
\par\end{centering}
\caption{The VEVs of the electric ($U=E$, full curves) and magnetic ($U=B$,
dashed curves) field squares (measured in units of $m^{D+1}$) as functions
of $z/a$ for the $D=3$ Proca field. The graphs are plotted for $%
ma=0.75,1,1.25,1.5$. The ratio $|\left\langle U^{2}\right\rangle |/m^{D+1}$
is a decreasing function of $ma$.}
\label{figE2}
\end{figure}

Figure \ref{figE2} displays the VEVs of the electric ($U=E$, full curves)
and magnetic ($U=B$, dashed curves) field squares (measured in units of $%
m^{D+1}$) for the Proca field in 3-dimensional space ($D=3$) versus $z/a$.
The graphs are plotted for $ma=0.75,1,1.25,1.5$. The VEVs $|\left\langle
U^{2}\right\rangle |/m^{D+1}$, with $U=E$ and $U=B$, are decreasing
functions of $ma$. 

The VEV of the electric field squared determines the Casimir-Polder
interaction energy between the plates and a polarizable particle. For an
isotropic polarizability, neglecting the effects of dispersion, the
potential is expressed as $U_{\mathrm{CP}}=-\alpha _{P}\left\langle
E^{2}\right\rangle /2$, where $\alpha _{P}$ is the static polarizability of
the particle. From the analysis of the electric field squared given above it
follows that for PMC conditions the Casimir-Polder force is repulsive with
respect to the nearest plate.

\begin{figure}[tbph]
\begin{centering}
\epsfig{figure=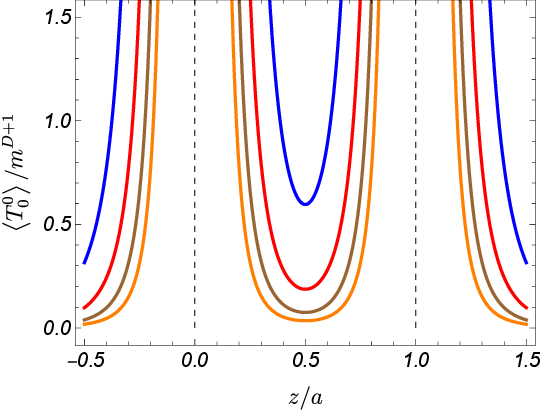,width=7.5cm,height=6.cm}
\par\end{centering}
\caption{Vacuum energy density (in units of $m^{D+1}$) versus $z/a$ for a
massive vector field in $D=3$ spatial dimensions. The graphs are plotted for 
$ma=0.75,1,1.25,1.5$. The ratio $\left\langle T_{0}^{0}\right\rangle
/m^{D+1} $ decreases with increasing $ma$.}
\label{figT00}
\end{figure}

In figure \ref{figT00} we have plotted the energy density for the Proca
field in 3-dimensional space ($D=3$) as a function of $z/a$ for $%
ma=0.75,1,1.25,1.5$. The ratio $\left\langle T_{0}^{0}\right\rangle /m^{D+1}$
decreases with increasing $ma$. 

\begin{figure}[tbph]
\begin{centering}
\epsfig{figure=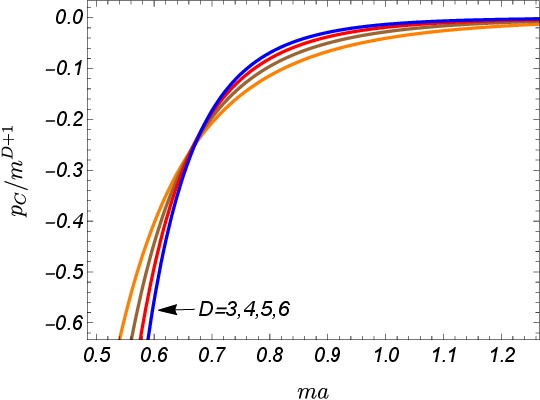,width=7.5cm,height=6.cm}
\par\end{centering}
\caption{The Casimir pressure for PMC conditions as a function of the
distance between the plates for $D=3,4,5,6$.}
\label{figpC}
\end{figure}

For small separations between the plates, compared with the Compton
wavelength, one has $ma\ll 1$. By using the expansion of the function $%
f_{\nu }(y)$ for small argument, given in the previous subsection, for the
leading term in the Casimir pressure we get%
\begin{equation}
p_{\mathrm{C}}\approx -\frac{D^{2}\Gamma \left( \frac{D+1}{2}\right) }{%
\left( 4\pi \right) ^{\frac{D+1}{2}}a^{D+1}}\zeta (D+1).  \label{pCnear}
\end{equation}%
This coincides with the result in the zero-mass limit. At large separations
between the plates, $ma\gg 1$, the Casimir forces are exponentially
suppressed:%
\begin{equation}
p_{\mathrm{C}}\approx -\frac{D\left( D+1\right) m^{D+1}}{\left( 2\pi \right)
^{\frac{D}{2}}\sqrt{2ma}}e^{-2ma}.  \label{pClarge}
\end{equation}%
Figure \ref{figpC} displays the Casimir pressure versus the distance between
the plates (in units of the Compton wavelength) for different values of the
spatial dimension $D$. The pressure is negative and corresponds to an
attractive Casimir force. 

\subsection{The Casimir energy}

We considered the local characteristics of the vacuum state. Away from
boundaries, their renormalization reduces to subtracting the parts
corresponding to the boundary-free geometry. A key global quantity in
discussions of the Casimir effect is the total vacuum energy, evaluated as
the sum of the zero-point energies of elementary oscillators. In the region
between the plates, $0\leq z\leq a$, the vacuum energy per unit surface area
of the plates is given by%
\begin{equation}
\mathcal{E}=\frac{1}{2}\int \frac{d\mathbf{k}}{\left( 2\pi \right) ^{D-1}}%
\,\left( \sum_{\sigma =1}^{D-1}\sum_{n=0}^{\infty }+\sum_{n=1}^{\infty
}\right) \omega _{n}(\mathbf{k}),  \label{Ecas}
\end{equation}%
where $\omega _{n}(\mathbf{k})=\sqrt{\mathbf{k}^{2}+(\pi n/a)^{2}+m^{2}}$
and the first and second sums in the braces correspond to the contributions
of the transverse and longitudinal polarizations, respectively. The
expression on the right-hand side of (\ref{Ecas}) is divergent and requires
regularization. Schemes for extracting the finite part of the vacuum energy
have been widely discussed in the theory of the Casimir effect \cite%
{Most97,Milt02,Bord09,Casi11}. Here, we briefly describe the zeta function
regularization approach (see also \cite{Eliz94,Kirs02}, for details and
applications).

We introduce related to (\ref{Ecas}) function%
\begin{equation}
\mathcal{E}(s)=\frac{1}{2}\int \frac{d\mathbf{k}}{\left( 2\pi \right) ^{D-1}}%
\,\left( \sum_{\sigma =1}^{D-1}\sum_{n=0}^{\infty }+\sum_{n=1}^{\infty
}\right) \omega _{n}^{-2s}(\mathbf{k}),  \label{Es}
\end{equation}%
and the zeta function%
\begin{equation}
\zeta (s)=\int \frac{d\mathbf{k}}{\left( 2\pi \right) ^{D-1}}%
\,\sum_{n=-\infty }^{+\infty }\frac{1}{\omega _{n}^{2s}(\mathbf{k})}.
\label{zetas}
\end{equation}%
These functions are finite in the range $\mathrm{Re}\,s>(D+1)/2$ of the complex variable $s$. The vacuum energy is obtained from (\ref{Es}) by analytic continuation at $s=-1/2$. A
convenient representation of the function (\ref{zetas}) can be found using
the Chowla-Selberg formula \cite{Eliz98}:%
\begin{equation}
\zeta (s)=\frac{am^{D-2s}}{2^{D-1}\pi ^{\frac{D}{2}}\Gamma (s)}\,\left[
\Gamma \left( s-\frac{D}{2}\right) +2^{\frac{D}{2}-s+2}\sum_{n=1}^{\infty
}f_{\frac{D}{2}-s}(2nma)\right] .  \label{zetas2}
\end{equation}%
With this representation, the function (\ref{Es}) is expressed in the form%
\begin{eqnarray}
\mathcal{E}(s) &=&\,\,\frac{Dam^{D-2s}\Gamma \left( s-\frac{D}{2}\right) }{%
2^{D+1}\pi ^{\frac{D}{2}}\Gamma (s)}+\left( D-2\right) \frac{%
m^{D-1-2s}\Gamma \left( s-\frac{D-1}{2}\right) }{2^{D+1}\pi ^{\frac{D-1}{2}%
}\Gamma (s)}  \notag \\
&&+\,\frac{Dam^{D-2s}}{2^{\frac{D}{2}+s-1}\pi ^{\frac{D}{2}}\Gamma (s)}%
\sum_{n=1}^{\infty }f_{\frac{D}{2}-s}(2nma).  \label{Es2}
\end{eqnarray}%
The dependence on the separation between the plates is contained in the last
term, which is finite at $s=-1/2$.

In the absence of boundaries, the function $\mathcal{E}_{0}(s)$
corresponding to the vacuum energy density is expressed as%
\begin{equation}
\mathcal{E}_{0}(s)=\frac{aD}{2}\int \frac{d\mathbf{k}_{D}}{\left( 2\pi
\right) ^{D}}\frac{1}{\left( \mathbf{k}_{D}^{2}+m^{2}\right) ^{2s}}\,,
\label{E0s}
\end{equation}%
where $D$ in front of the integral stands for the number of polarizations.
By evaluating the integral, we can show that $\mathcal{E}_{0}(s)$ coincides
with the first term in the right-hand side of (\ref{Es2}). Therefore, the
change in the vacuum energy caused by the presence of boundaries (the
Casimir energy) is given by 
\begin{equation}
\mathcal{E}_{\mathrm{C}}=\left[ \mathcal{E}(s)-\mathcal{E}_{0}(s)\right]
|_{s=-1/2},  \label{Ecas2}
\end{equation}%
where $|_{s=-1/2}$ is understood in the sense of analytic continuation. From
(\ref{Es2}) we obtain 
\begin{equation}
\mathcal{E}_{\mathrm{C}}=\frac{\left( D-2\right) m^{D-1}}{2^{D+1}\pi ^{\frac{%
D-1}{2}}}\left. \frac{\Gamma \left( s-\frac{D-1}{2}\right) }{m^{2s}\Gamma (s)%
}\right\vert _{s=-1/2}-\,\frac{2Dam^{D+1}}{\left( 2\pi \right) ^{\frac{D+1}{2%
}}}\sum_{n=1}^{\infty }f_{\frac{D+1}{2}}(2nma).  \label{Ecas3}
\end{equation}%
The first term on the right-hand side is independent of $a$ and can be
absorbed through the renormalization of the self-energies of the boundaries.
Now, by using (\ref{relfnu}), we get the relation $p_{\mathrm{C}}=-\partial 
\mathcal{E}_{\mathrm{C}}/\partial a$ between the Casimir energy and
pressure. Note that the same $a$-dependent part of the Casimir energy can be
obtained by applying the Abel-Plana formula (\ref{APF}) to the series over $%
n $ in (\ref{Ecas}). The contribution of the integral in (\ref{APF}) gives
the vacuum energy in the absence of boundaries.

Based on the above results, the vacuum expectation values can also be obtained
in the Stueckelberg formulation for a massive vector field (see \cite{Rueg04}
for a review). The action in Stueckelberg electromagnetism is obtained from
the Proca action (\ref{S}) by replacing 
\begin{equation}
A_{\mu }=A_{\mathrm{(S)}\mu }+2\frac{\sqrt{\pi }}{m}\partial _{\mu }\varphi ,
\label{AtoAS}
\end{equation}%
where $A_{\mathrm{(S)}\mu }$ is the Stueckelberg vector field and $\varphi $
is an auxiliary scalar field. The Lagrangian density is separated into two
parts. The first part corresponds to the Lagrangian density for the field $%
A_{\mathrm{(S)}\mu }$ given by (\ref{S}) with the replacement $A_{\mu
}\rightarrow A_{\mathrm{(S)}\mu }$. The second contribution corresponds to a
massless scalar field $\varphi $ interacting with the vector field through
the term $mA^{\mu }\partial _{\mu }\varphi /(2\sqrt{\pi })$ in the
Lagrangian density. The total action is invariant under the gauge
transformation%
\begin{equation}
A_{\mathrm{(S)}\mu }^{\prime }=A_{\mathrm{(S)}\mu }+\partial _{\mu }\psi
,\;\varphi ^{\prime }=\varphi -\frac{m\psi }{2\sqrt{\pi }},  \label{ASgauge}
\end{equation}%
where the function $\psi $ obeys the equation $\left( \partial _{\mu
}\partial ^{\mu }+m^{2}\right) \psi =0$. In particular, $F_{\mu \nu
}(A_{\alpha })=F_{\mu \nu }(A_{\mathrm{(S)}\alpha })$ and the boundary
condition $n^{\nu }F_{\mu \nu }=0$ is the same for both the Proca and
Stueckelberg fields. In the canonical quantization procedure, a simple
gauge-fixing choice corresponds to the Stueckelberg (unitary) gauge with
a zero Stueckelberg scalar, $\varphi =0$. This reduces the action to the
Proca action for the field $A_{\mathrm{(S)}\mu }$ and the corresponding
analysis follows the steps described above. 

In the covariant quantization
procedure, a gauge-fixing term $-( \partial _{\mu }A_{\mathrm{(S)}%
}^{\mu }-2\sqrt{\pi }\xi m\varphi ) ^{2}/(8\pi \xi )$ is added to the
Stueckelberg Lagrangian density for the fields $A_{\mathrm{(S)}\mu }$ and $%
\varphi $, with a real parameter $\xi $. In the special case of the
Stueckelberg-Feynman gauge, corresponding to $\xi =-1$, the interaction term
in the initial action cancels out with the part coming from the gauge-fixing
term. Up to total derivatives, the Lagrangian density decomposes into vector
and scalar contributions with a free scalar field $\varphi $ of mass $m$.
The detailed analysis of the corresponding model in general curved
backgrounds is presented in \cite{Belo16}. As an application, the Casimir
energy-momentum tensor is considered in Stueckelberg electromagnetism for a
reflecting planar boundary on a 4-dimensional Minkowski spacetime
background. It has been shown that the total energy-momentum tensor,
including the contributions from the vector and scalar fields, coincides
with the result obtained in \cite{Davi81} for the Proca field. The
corresponding expression is a special case of (\ref{Tll1}) for $D=3$. In 
\cite{Belo16}, the separate contributions of the fields $A_{\mathrm{(S)}\mu }
$ and $\varphi $ to the vacuum energy-momentum tensor are also discussed.
The part corresponding to the vector field reduces to the corresponding
result for Maxwell's field in the zero-mass limit.

\section{Two-point functions and the Casimir densities for a massless vector
field with PMC conditions}

\label{sec:Massless}

In order to compare the zero-mass limit of the VEVs for Proca field with PMC
boundary conditions, in this section we consider the two-point functions and
the VEVs for a massless vector field. The total vacuum energy for PMC
Casimir piston in $(D+1)$-dimensional spacetime has been investigated in 
\cite{Eder08b}.

\subsection{Modes and the two-point functions for the field tensor}

The action for a massless vector field is invariant under the gauge
transformation $A_{\mu }^{\prime }(x)=A_{\mu }(x)+\partial _{\mu }\chi (x)$.
Fixing the gauge by the conditions $\partial _{\mu }A^{\mu }(x)=0$ and $%
A_{D}(x)=0$, for the mode functions with $k_{D}\neq 0$ we have%
\begin{equation}
A_{(\beta )l}^{(0)}=C_{\beta }\epsilon _{(\sigma )l}^{(0)}\cos \left(
k^{D}z\right) e^{-ik_{j}x_{\parallel }^{j}},\;A_{(\beta )D}^{(0)}=0.
\label{Abetm0}
\end{equation}%
where%
\begin{equation}
k^{D}=\frac{\pi n}{a},\;k^{0}=\omega =\sqrt{|\mathbf{k}|^{2}+(k^{D})^{2}}%
,\;n=1,2,\ldots .  \label{kDm0}
\end{equation}
In this case the longitudinal polarization is absent and $\sigma =1,\ldots
,D-1$. As before, one has $k^{l}\epsilon _{(\sigma )l}^{(0)}=0$, $%
g^{il}\epsilon _{(\sigma ^{\prime })i}^{(0)}\epsilon _{(\sigma
)l}^{(0)}=-\delta _{\sigma ^{\prime }\sigma }$, and the relation%
\begin{equation}
\sum_{\sigma =1}^{D-1}\epsilon _{(\sigma )i}^{(0)}\epsilon _{(\sigma
)l}^{(0)}=\frac{k_{i}k_{l}}{k_{D}^{2}}-g_{il}.  \label{epsm0}
\end{equation}%
The normalization coefficient $C_{\beta }$ in (\ref{Abetm0}) is given by (%
\ref{Cbet1}) with $n\neq 0$ and $\omega $ from (\ref{kDm0}).

In addition, there are also zero modes with 
\begin{equation}
A_{(0\beta )l}^{(0)}=C_{\beta }\epsilon _{(0\sigma
)l}^{(0)}e^{-ik_{j}x_{\parallel }^{j}},\;A_{(0\beta )D}^{(0)}=0,
\label{Abetm00}
\end{equation}%
where $g_{il}k^{i}k^{l}=0$ and $k^{0}=\omega =|\mathbf{k}|$. In this case we
have $D-2$ polarization degrees of freedom ($\sigma =1,2,\ldots ,D-2$) with $%
\epsilon _{(0\sigma )0}^{(0)}=0$. The polarization vector obeys the relations%
\begin{equation}
\epsilon _{(0\sigma )l}^{(0)}\epsilon _{(0\sigma ^{\prime })}^{(0)l}=-\delta
_{\sigma \sigma ^{\prime }},\;\sum_{\sigma =1}^{D-2}\epsilon _{(0\sigma
)i}^{(0)}\epsilon _{(0\sigma )l}^{(0)}=-\frac{k_{i}k_{l}}{|\mathbf{k}|^{2}}%
-g_{il},  \label{eps0rel}
\end{equation}%
where $i,l=1,2,\ldots ,D-1$. The coefficient $C_{\beta }$ in (\ref{Abetm00})
is obtained from (\ref{Cbet1}) with $n=0$ and $\omega =|\mathbf{k}|$.

By using the mode functions (\ref{Abetm0}) and (\ref{Abetm00}), for the
two-point functions of the field tensor in the region between the plates we
get 
\begin{align}
\left\langle F_{\mu \nu }F_{\rho \sigma }^{\prime }\right\rangle ^{(0)}&
=2\left( \eta _{\lbrack \mu \sigma }\partial _{\nu ]}\partial _{\rho
}^{\prime }-\eta _{\lbrack \mu \rho }\partial _{\nu ]}\partial _{\sigma
}^{\prime }\right) A_{0}(x,x^{\prime }),  \notag \\
\left\langle F_{D\mu }F_{D\nu }^{\prime }\right\rangle ^{(0)}& =\partial
_{\mu }\partial _{\nu }^{\prime }B_{0}(x,x^{\prime })-\eta _{\mu \nu
}\partial _{D}\partial _{D}^{\prime }A_{0}(x,x^{\prime }),  \label{FFm0b}
\end{align}%
where the functions $A_{0}(x,x^{\prime })$ and $B_{0}(x,x^{\prime })$ are
defined by (\ref{Bxx}) with $m=0$. We note that the contribution of the term 
$n=0$ in $A_{0}(x,x^{\prime })$ comes from the zero modes (\ref{Abetm00}).
Simpler expressions for the functions $A_{0}(x,x^{\prime })$ and $%
B_{0}(x,x^{\prime })$ are obtained from (\ref{Bxx2}) in the limit $%
m\rightarrow 0$:%
\begin{align}
A_{0}(x,x^{\prime })& =\frac{\Gamma \left( \frac{D-1}{2}\right) }{\pi ^{%
\frac{D-1}{2}}}\sum_{j=\mp 1}\sum_{n=-\infty }^{+\infty }\frac{1}{%
b_{j,n}^{D-1}(x,x^{\prime })},  \notag \\
B_{0}(x,x^{\prime })& =-\frac{\Gamma \left( \frac{D-1}{2}\right) }{\pi ^{%
\frac{D-1}{2}}}\sum_{j=\mp 1}\sum_{n=-\infty }^{+\infty }\frac{j}{%
b_{j,n}^{D-1}(x,x^{\prime })}.  \label{B0xx2}
\end{align}%
Evaluating the derivatives, we get%
\begin{equation}
\left\langle F_{\mu \nu }F_{\rho \sigma }^{\prime }\right\rangle
^{(0)}=\left\langle F_{\mu \nu }F_{\rho \sigma }^{\prime }\right\rangle
_{m\rightarrow 0},  \label{FF0lim}
\end{equation}%
and the two-point function of the field tensor for a massless vector field
coincides with the zero-mass limit of the corresponding function for a
massive vector field, given by (\ref{FFm0lim}).

\subsection{VEVs and the Casimir forces}

Given the two-point functions for the field tensor we can evaluate the VEVs
of physical observables. By taking into account the relation (\ref{FF0lim}),
we conclude that the VEVs of the electric and magnetic field squares and the
condensate coincide with the corresponding VEVs of the massive field in the
zero-mass limit:%
\begin{equation}
\left\langle E^{2}\right\rangle ^{(0)}=\left\langle E^{2}\right\rangle
_{m\rightarrow 0},\;\left\langle B^{2}\right\rangle ^{(0)}=\left\langle
B^{2}\right\rangle _{m\rightarrow 0},  \label{EBm0EB}
\end{equation}%
and $\left\langle F_{\mu \nu }F^{\mu \nu }\right\rangle ^{(0)}=\left\langle
F_{\mu \nu }F^{\mu \nu }\right\rangle _{m\rightarrow 0}$. The corresponding
expressions are given by (\ref{B2m0}) and (\ref{FFm0}). By using the mode
functions (\ref{Abetm0}) and (\ref{Abetm00}), for the nonzero components of
the vacuum energy-momentum tensor we find (no summation over $l$)%
\begin{align}
\left\langle T_{l}^{l}\right\rangle ^{(0)}& =\frac{\left( 1-D\right) \Gamma
\left( \frac{D+1}{2}\right) }{\left( 4\pi \right) ^{\frac{D+1}{2}}a^{D+1}}%
\left[ \zeta (D+1)-\sum_{n=-\infty }^{+\infty }\frac{\left( D-3\right) /2}{%
\left\vert n-z/a\right\vert ^{D+1}}\right] ,  \notag \\
\left\langle T_{D}^{D}\right\rangle ^{(0)}& =\frac{D\left( D-1\right) \Gamma
\left( \frac{D+1}{2}\right) }{\left( 4\pi \right) ^{\frac{D+1}{2}}a^{D+1}}%
\zeta (D+1),  \label{TDDm0}
\end{align}%
with $l=0,1,\ldots ,D-1$. For $D=3$ the massless vector field is conformally
invariant and the vacuum energy-momentum tensor is traceless: $\left\langle
T_{\mu }^{\mu }\right\rangle ^{(0)}=0$. Comparing the VEVs (\ref{TDDml0})
and (\ref{TDDm0}), we see that they are different: $\left\langle T_{\mu \nu
}\right\rangle _{m\rightarrow 0}\neq \left\langle T_{\mu \nu }\right\rangle
^{(0)}$. The Casimir densities for a massless vector field in the geometry
of two parallel plates on the background of $(D+1)$-dimensional anti-de
Sitter (AdS) spacetime (the plates are parallel to the AdS boundary) were
investigated in \cite{Saha20} for both PMC and PEC boundary conditions (for
the electromagnetic Casimir densities in the geometry of planar boundaries
in de Sitter spacetime see \cite{Saha14,Kota15}). As it has been shown in 
\cite{Saha20}, the VEVs $\left\langle E^{2}\right\rangle ^{(0)}$, $%
\left\langle B^{2}\right\rangle ^{(0)}$, and $\left\langle T_{\mu }^{\nu
}\right\rangle ^{(0)}$ are obtained from the results for the AdS bulk in the
limit when the curvature radius goes to infinity.

For a massive field, the VEV of the energy-momentum tensor, in addition to
the terms with the expectation values of the products of the field tensor,
contains the terms with the products of the vector potential multiplied by
the mass squared. As it has been discussed above, the zero-mass limit of
those parts does not vanish. As a consequence, the VEV of the
energy-momentum tensor for a massless field differs from the corresponding
VEV for a massive field in the zero-mass limit. We have the following
relation 
\begin{equation}
\left\langle T_{\mu \nu }\right\rangle _{m\rightarrow 0}=\left\langle T_{\mu
\nu }\right\rangle ^{(0)}+\lim_{m\rightarrow 0}\frac{m^{2}}{4\pi }\left(
\left\langle A_{(\mu }A_{\nu )}\right\rangle -\frac{1}{2}g_{\mu \nu
}\left\langle A_{\alpha }A^{\alpha }\right\rangle \right) ,  \label{Tdif}
\end{equation}%
for the corresponding VEVs. This relation is easy to check by using the
expression (\ref{AA3m0}) for the zero-mass limit of the VEVs biliniear in
the vector field.

\section{The Casimir densities for PEC conditions}

\label{sec:BCcond}

In this section, we consider the two-point functions and the Casimir
densities for the boundary condition being the higher dimensional
generalization of the condition imposed on the surface of perfect conductor
in spatial dimensions $D=3$ (PEC boundary condition). In terms of the dual
tensor $^{\ast }F_{\mu _{1}\cdots \mu _{D-1}}=\varepsilon _{\mu \nu \mu
_{1}\cdots \mu _{D-1}}F^{\mu \nu }/(D-1)!$, with $\varepsilon _{\mu \nu \mu
_{1}\cdots \mu _{D-1}}$ being the Levi-Civita tensor, the boundary
conditions are written as%
\begin{equation}
n^{\mu _{1}}\,^{\ast }F_{\mu _{1}\cdots \mu _{D-1}}=0,\;z=0,a.  \label{BCc}
\end{equation}%
In the geometry under consideration they are reduced to $F_{il}=0$ for $%
z=0,a $, and $i,l=0,1,\ldots ,D-1$.

\subsection{Mode functions}

Similar to the case of PMC boundary conditions, we have $D-1$ transverse
polarizations ($\sigma =1,2,\ldots ,D-1$) and a longitudinal polarization ($%
\sigma =D$). For the longitudinal polarization one has $F_{il}=0$ and the
conditions (\ref{BCc}) impose no restriction on the eigenvalues of the
momentum $k^{D}$ for that mode. For polarizations $\sigma =1,2,\ldots ,D-1$
the mode functions of the vector potential in the region $0\leq z\leq a$ are
presented in the form%
\begin{equation}
A_{(\beta )l}(x)=C_{\beta }\epsilon _{(\sigma )l}\sin \left( k^{D}z\right)
e^{-ik_{j}x_{\parallel }^{j}},\;A_{(\beta )D}=0,  \label{Atrc}
\end{equation}%
with the eigenvalues of the momentum $k^{D}$ determined by the boundary
condition at $z=a$:%
\begin{equation}
k^{D}=\frac{\pi n}{a},\;n=1,2,\ldots  \label{kDc}
\end{equation}%
The normalization constant is determined from the condition (\ref{NC2}) and
coincides with (\ref{Cbet1}). Note that in this case $n\neq 0$ and the zero
mode is absent. The mode functions for the longitudinal polarization ($%
\sigma =D$) are given by 
\begin{equation}
A_{(\beta )l}=-\frac{k_{l}k^{D}}{\lambda ^{2}}C_{\beta }e^{-ik_{\mu }x^{\mu
}},\;A_{(\beta )D}=C_{\beta }e^{-ik_{\mu }x^{\mu }}.  \label{Alc}
\end{equation}%
For these modes the components $F_{(\beta )il}$ of the field tensor vanish
and they obey the boundary condition for $-\infty <k^{D}<+\infty $. The
longitudinal mode is free to propagate in the region $-\infty <z<+\infty $
and in the normalization condition (\ref{NC2}) the integration over $z$ goes
over that region. For the normalization constant we get%
\begin{equation}
\left\vert C_{\beta }\right\vert ^{2}=\frac{\lambda ^{2}}{\left( 2\pi
\right) ^{D-1}m^{2}\omega },\;\sigma =D.  \label{Cbetc}
\end{equation}%
The mode functions (\ref{Alc}) coincide with the corresponding modes in the
boundary-free problem with $-\infty <z<+\infty $. The Casimir contributions
to the two-point functions and VEVs are obtained by subtracting from the
total VEVs the parts corresponding to the boundary-free geometry. Thus, the
longitudinal modes $\sigma =D$ will not contribute to the Casimir parts and
in the discussion below we present the contributions in the expectation
values corresponding to the transverse modes (polarizations $\sigma
=1,2,\ldots ,D-1$).

\subsection{Two-point functions}

The contribution of the transverse modes to the expectation values will be
denoted by $\left\langle \cdots \right\rangle _{\mathrm{tr}}$. For the
components of the two-point function $\left\langle A_{\mu }A_{\nu }^{\prime
}\right\rangle _{\mathrm{tr}}$ with $\mu =D$ or $\nu =D$ we have $%
\left\langle A_{D}A_{\nu }^{\prime }\right\rangle _{\mathrm{tr}%
}=\left\langle A_{\mu }A_{D}^{\prime }\right\rangle _{\mathrm{tr}}=0$. For
the evaluation of the VEVs of the fields squared and energy-momentum tensor
one needs the expectation values $\left\langle A_{l}A_{p}^{\prime
}\right\rangle _{\mathrm{tr}}$ in the coincidence limit $x^{\prime
}\rightarrow x$. In that limit they are diagonal and from the problem
symmetry $\left\langle A_{0}A^{0}\right\rangle _{\mathrm{tr}}=\cdots
=\left\langle A_{D-1}A^{D-1}\right\rangle _{\mathrm{tr}}$. Hence, it
suffices to consider the two-point function $\left\langle A_{l}A^{l\prime
}\right\rangle _{\mathrm{tr}}$. By using the modes (\ref{Atrc}) one obtains%
\begin{equation}
\left\langle A_{l}A^{l\prime }\right\rangle _{\mathrm{tr}}=(1-D)B(x,x^{%
\prime }),  \label{All}
\end{equation}%
with the function $B(x,x^{\prime })$ defined by (\ref{Bxx}) and presented in
the form (\ref{Bxx2}).

The two-point functions of the field tensor are also expressed in terms of
the function $B(x,x^{\prime })$:%
\begin{align}
\left\langle F_{ik}F_{lp}^{\prime }\right\rangle _{\mathrm{tr}}& =2\left(
g_{[ip}\partial _{k]}\partial _{l}^{\prime }-g_{[il}\partial _{k]}\partial
_{p}^{\prime }\right) B(x,x^{\prime }),  \notag \\
\left\langle F_{Dk}F_{lp}^{\prime }\right\rangle _{\mathrm{tr}}&
=2g_{k[l}\partial _{p]}^{\prime }\partial _{D}B(x,x^{\prime }),  \notag \\
\left\langle F_{Dl}F_{D}^{\prime \cdot l}\right\rangle _{\mathrm{tr}}&
=\left( 1-D\right) \partial _{D}\partial _{D}^{\prime }B(x,x^{\prime }).
\label{FFtr}
\end{align}%
By using the expression (\ref{Bxx2}) we get the representation%
\begin{equation}
\left\langle F_{\mu \nu }F_{\rho \sigma }^{\prime }\right\rangle _{\mathrm{tr%
}}=\frac{4m^{D+1}}{\left( 2\pi \right) ^{\frac{D-1}{2}}}\sum_{j=\mp
1}\sum_{n=-\infty }^{\infty }jG_{\mu \nu \rho \sigma }^{(j,n)}(x,x^{\prime
}),  \label{FF2ptr}
\end{equation}%
where $G_{iklp}^{(j,n)}(x,x^{\prime })=F_{iklp}^{(j,n)}(x,x^{\prime })$ and $%
G_{Dklp}^{(j,n)}(x,x^{\prime })=F_{Dklp}^{(j,n)}(x,x^{\prime })$ with the
components $F_{\mu \nu \rho \sigma }^{(j,n)}(x,x^{\prime })$ from (\ref{FF2p}%
). The components with $\mu =\rho =D$, required in the evaluation of the
VEVs in the coincidence limit are obtained form the two-point function $%
\left\langle F_{Dl}F_{D}^{\cdot l\prime }\right\rangle _{\mathrm{tr}}$ and
for it we have%
\begin{equation}
G_{DlD}^{(j,n)\cdot \cdot \cdot l}(x,x^{\prime })=j\frac{D-1}{2}\left[
m^{2}\left( 2na-z_{j}\right) ^{2}f_{\frac{D+3}{2}}(mb_{j,n}(x,x^{\prime
}))-f_{\frac{D+1}{2}}(mb_{j,n}(x,x^{\prime }))\right] .  \label{GDl}
\end{equation}

\subsection{The Casimir densities}

The expectation values of the physical characteristics for the vacuum state
are obtained from the two-point functions in the limit $x^{\prime
}\rightarrow x$. For points away from boundaries the renormalization is
reduced to the subtraction of the part in the two-point functions with $n=0$
and $j=-1$. That part presents the contribution corresponding to the
geometry without boundaries. As it has been discussed above, the
longitudinal mode does not contribute to the subtracted (renormalized) VEVs
and they will be denoted by $\left\langle \cdots \right\rangle $ (omitting
the subscript tr).

We start with the VEV $\left\langle A_{\mu }A_{\nu }\right\rangle $. One has 
$\left\langle A_{D}A_{\nu }\right\rangle =0$ and 
\begin{equation}
\left\langle A_{i}A_{l}\right\rangle =2g_{il}\frac{\left( D-1\right) m^{D-1}%
}{\left( 2\pi \right) ^{\frac{D-1}{2}}D}\left[ -2\sum_{n=1}^{\infty }f_{%
\frac{D-1}{2}}(2nma)+\sum_{n=-\infty }^{+\infty }f_{\frac{D-1}{2}}(2mz_{n})%
\right] .  \label{Ailc}
\end{equation}%
Note that unlike to the case of the boundary condition (\ref{BC}), the VEV $%
\left\langle A_{i}A_{l}\right\rangle $ is finite in the zero-mass limit.
Next we consider the condensate $\left\langle F_{\mu \nu }F^{\mu \nu
}\right\rangle $. From (\ref{FF2ptr}) one obtains%
\begin{align}
\left\langle F_{\mu \nu }F^{\mu \nu }\right\rangle & =\frac{4(1-D)m^{D+1}}{%
\left( 2\pi \right) ^{\frac{D-1}{2}}}\left\{ 2\sum_{n=1}^{\infty }f_{\frac{%
D-1}{2}}(2nma)\right.  \notag \\
& \left. +\sum_{n=-\infty }^{\infty }\left[ 2Df_{\frac{D+1}{2}}(2mz_{n})+f_{%
\frac{D-1}{2}}(2mz_{n})\right] \right\} .  \label{FFr2c}
\end{align}%
The condensate is negative for $D\geq 2$ and vanishes for $D=1$. Combining
this with the expectation value (\ref{Ailc}), we get the VEV of the
Lagrangian density:%
\begin{equation}
\left\langle L(A_{\mu })\right\rangle =\frac{(D-1)m^{D+1}}{\left( 2\pi
\right) ^{\frac{D+1}{2}}}\sum_{n=-\infty }^{\infty }\left[ Df_{\frac{D+1}{2}%
}(2mz_{n})+f_{\frac{D-1}{2}}(2mz_{n})\right] .  \label{Lc}
\end{equation}%
This VEV will be used in the evaluation of the energy-momentum tensor.

The VEVs of the electric and magnetic field squares are obtained by using
the formulas (\ref{E2}) and (\ref{B20}). They are given by the expressions%
\begin{align}
\left\langle E^{2}\right\rangle & =\frac{2\left( D-1\right) m^{D+1}}{\left(
2\pi \right) ^{\frac{D-1}{2}}D}\left\{ 2\sum_{n=1}^{\infty }\left[ f_{\frac{%
D-1}{2}}(2nma)-Df_{\frac{D+1}{2}}(2nma)\right] \right.  \notag \\
& \left. +\sum_{n=-\infty }^{\infty }\left[ 3Df_{\frac{D+1}{2}}(2mz_{n})+f_{%
\frac{D-1}{2}}(2mz_{n})\right] \right\} ,  \label{E2c}
\end{align}%
and%
\begin{align}
\left\langle B^{2}\right\rangle & =\frac{2(1-D)m^{D+1}}{\left( 2\pi \right)
^{\frac{D-1}{2}}}\left\{ 2\sum_{n=1}^{\infty }\left[ f_{\frac{D+1}{2}}(2nma)+%
\frac{D-1}{D}f_{\frac{D-1}{2}}(2nma)\right] \right.  \notag \\
& \left. +\sum_{n=-\infty }^{\infty }\left[ \left( 2D-3\right) f_{\frac{D+1}{%
2}}(2mz_{n})+\frac{D-1}{D}f_{\frac{D-1}{2}}(2mz_{n})\right] \right\} .
\label{B2c}
\end{align}%
From the inequality (\ref{Ineq}) it follows that the VEV $\left\langle
E^{2}\right\rangle $ is positive for $D\geq 2$. The VEV of the magnetic
field squared vanishes for $D=1$ and is negative in spatial dimensions $%
D\geq 2$.

The expressions for the energy density and parallel stresses read (no
summation over $l$) 
\begin{align}
\left\langle T_{l}^{l}\right\rangle & =\frac{\left( 1-D\right) m^{D+1}}{%
\left( 2\pi \right) ^{\frac{D+1}{2}}}\left\{ 2\sum_{n=1}^{\infty }f_{\frac{%
D+1}{2}}(2nma)\right.  \notag \\
& \left. +\sum_{n=-\infty }^{\infty }\left[ \left( D-3\right) f_{\frac{D+1}{2%
}}(2mz_{n})+\frac{D-2}{D}f_{\frac{D-1}{2}}(2mz_{n})\right] \right\} .
\label{Tilc}
\end{align}%
The vacuum energy density is negative for $D\geq 3$ and positive for $D=2$.
The normal stress is presented as 
\begin{equation}
\left\langle T_{D}^{D}\right\rangle =\frac{2\left( D-1\right) m^{D+1}}{%
\left( 2\pi \right) ^{\frac{D+1}{2}}}\sum_{n=1}^{\infty }\left[ Df_{\frac{D+1%
}{2}}(2nma)+f_{\frac{D-1}{2}}(2nma)\right] .  \label{TDc}
\end{equation}%
The VEV of the energy-momentum tensor vanishes for $D=1$. As expected from
the conservation equation for the energy-momentum tensor, the distribution
of the normal stress in the region between the plates is uniform. For the
Casimir pressure on the plates we have $p_{\mathrm{C}}=-\left\langle
T_{D}^{D}\right\rangle $ and the Casimir forces are attractive. Note that we
have a simple relation between the Casimir forces for PMC and PEC boundary
conditions:%
\begin{equation}
\frac{\left\langle T_{D}^{D}\right\rangle _{\mathrm{PMC}}}{\left\langle
T_{D}^{D}\right\rangle _{\mathrm{PEC}}}=\frac{D}{D-1}.  \label{TDDratio}
\end{equation}%
Note that the expression on the right-hand side of (\ref{TDDratio}) is the
ratio of the number of independent polarizations of the vector field
influenced by PMC and PEC boundary conditions. For the trace of the
energy-momentum tensor we obtain%
\begin{align}
\left\langle T_{l}^{l}\right\rangle & =\frac{\left( 1-D\right) m^{D+1}}{%
\left( 2\pi \right) ^{\frac{D+1}{2}}}\left\{ -2\sum_{n=1}^{\infty }f_{\frac{%
D-1}{2}}(2nma)\right.  \notag \\
& \left. +\sum_{n=-\infty }^{\infty }\left[ D\left( D-3\right) f_{\frac{D+1}{%
2}}(2mz_{n})+\left( D-2\right) f_{\frac{D-1}{2}}(2mz_{n})\right] \right\} .
\label{TraceC}
\end{align}%
For $D=3$ it vanishes in the massless limit and the trace anomaly for PEC
conditions is absent.

The vacuum densities in the problem with a single plate $z=0$ are given by
the terms $n=0$ in the expressions given above. For the VEVs of the electric
and magnetic field squares one has%
\begin{align}
\left\langle E^{2}\right\rangle _{1}& =\frac{2\left( D-1\right) m^{D+1}}{%
\left( 2\pi \right) ^{\frac{D-1}{2}}D}\left[ 3Df_{\frac{D+1}{2}}(2m|z|)+f_{%
\frac{D-1}{2}}(2m|z|)\right] ,  \notag \\
\left\langle B^{2}\right\rangle _{1}& =\frac{2(1-D)m^{D+1}}{\left( 2\pi
\right) ^{\frac{D-1}{2}}}\left[ \left( 2D-3\right) f_{\frac{D+1}{2}}(2m|z|)+%
\frac{D-1}{D}f_{\frac{D-1}{2}}(2m|z|)\right] .  \label{E2B2pl1}
\end{align}%
They vanish for $D=1$. The VEV of the electric field squared is positive and
the VEV of the magnetic field squared is negative for $D\geq 2$. The vacuum
energy density and the parallel stresses are expressed as (no summation over 
$l$) 
\begin{equation}
\left\langle T_{l}^{l}\right\rangle _{1}=\frac{\left( 1-D\right) m^{D+1}}{%
\left( 2\pi \right) ^{\frac{D+1}{2}}}\left[ \left( D-3\right) f_{\frac{D+1}{2%
}}(2m|z|)+\frac{D-2}{D}f_{\frac{D-1}{2}}(2m|z|)\right] ,  \label{Tll1c}
\end{equation}%
and the normal stress vanishes. As seen, the Casimir densities for the PEC
conditions vanish for spatial dimension $D=1$. We could expect that result,
by taking into account that for $D=1$ the only polarization mode is
longitudinal and the PEC conditions do not affect that mode.

Considering the Casimir energy in the region between the plates follows the
same lines of reasoning as the PMC condition described above. The
corresponding expression is as follows:%
\begin{equation}
\mathcal{E}_{\mathrm{C}}=\frac{\left( 1-D\right) m^{D-1}}{2^{D+1}\pi ^{\frac{%
D-1}{2}}}\left. \frac{\Gamma \left( s-\frac{D-1}{2}\right) }{m^{2s}\Gamma (s)%
}\right\vert _{s=-1/2}-\left( D-1\right) \frac{2am^{D+1}}{\left( 2\pi
\right) ^{\frac{D+1}{2}}}\sum_{n=1}^{\infty }f_{\frac{D+1}{2}}(2nma).
\label{EcasPEC}
\end{equation}%
The coefficient $(D-1)$ in the last term shows that only the transverse
polarizations contribute to the Casimir energy. In this case, the relation $%
\partial \mathcal{E}_{\mathrm{C}}/\partial a=\left\langle
T_{D}^{D}\right\rangle =-p_{\mathrm{C}}$ also holds.

\begin{figure}[tbph]
\begin{centering}
\epsfig{figure=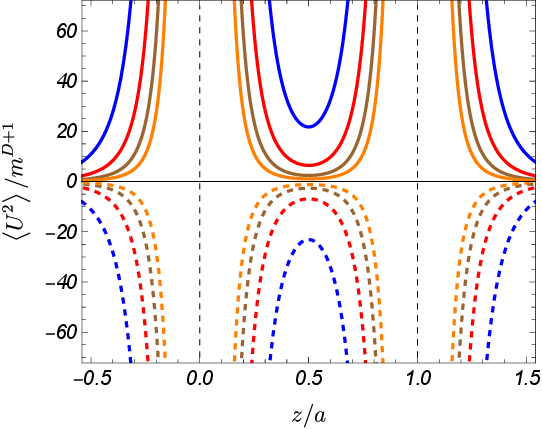,width=7.5cm,height=6.cm}
\par\end{centering}
\caption{VEVs of the electric and magnetic field squares (in units $m^{D+1}$%
) for the Proca field with PEC boundary conditions in 3-dimensional space as
functions of $z/a$. For the parameter $ma$ the values $ma=0.75,1,1.25,1.5$
are taken (the ratio $|\left\langle U^{2}\right\rangle |/m^{D+1}$ is a
decreasing function of $ma$). }
\label{figE2c}
\end{figure}

\begin{figure}[tbph]
\begin{centering}
\epsfig{figure=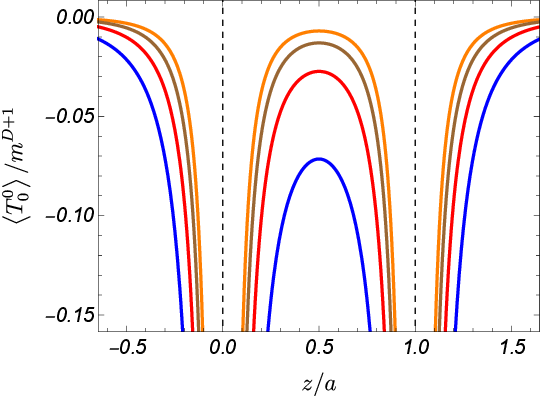,width=7.5cm,height=6.cm}
\par\end{centering}
\caption{Vacuum energy density (in units of $m^{D+1}$) for a $D=3$ massive
vector field with PEC conditions versus $z/a$. The graphs are plotted for
the values $ma=0.75,1,1.25,1.5$. The modulus of the ratio $\left\langle
T_{0}^{0}\right\rangle /m^{D+1}$ decreases with increasing $ma$.}
\label{figT00c}
\end{figure}

Figure \ref{figE2c} presents the dependence of the electric ($U=E$, full
curves) and magnetic ($U=B$, dashed curves) field squares (in units $m^{D+1}$%
) on $z/a$. The graphs are plotted for $ma=0.75,1,1.25,1.5$ (the ratio $%
|\left\langle U^{2}\right\rangle |/m^{D+1}$ decreases with increasing $m$).
The analysis for the VEV $\left\langle E^{2}\right\rangle $ given above
shows that for PEC boundary conditions the Casimir-Polder force acting on a
polarizable microparticle is attractive with respect to the nearest plate
for spatial dimensions $D\geq 2$. 
The vacuum energy density, measured in units $m^{D+1}$ and given by (\ref%
{Tilc}) and (\ref{Tll1c}) with $l=0$, is plotted in figure \ref{figT00c} as
a function of $z/a$ for the same values $ma=0.75,1,1.25,1.5$. The modulus of
the ratio $\left\langle T_{0}^{0}\right\rangle /m^{D+1}$ decreases with
increasing $ma$. 
The behavior of the Casimir pressure $p_{\mathrm{C}}$ for a vector field
with PEC conditions on the plates, as a function of $ma$, directly follows
from figure \ref{figpC} by taking into account the relation (\ref{TDDratio}).

\begin{figure}[tbph]
\begin{centering}
\epsfig{figure=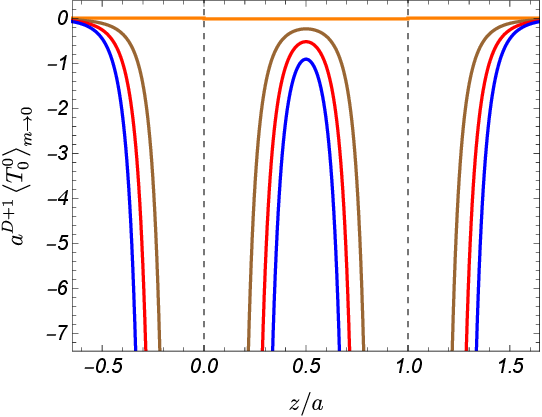,width=7.5cm,height=6.cm}
\par\end{centering}
\caption{Vacuum energy density (in units $1/a^{D+1}$) as a function of $z/a$
for a massless vector field with PEC boundary conditions. The graphs are
plotted for the values of the spatial dimension $D=3,4,5,6$. The product $%
a^{D+1}|\left\langle T_{0}^{0}\right\rangle |$ increases with increasing $D$%
. }
\label{figT00cm0}
\end{figure}

In the massless limit the expressions for the VEVs of the electric and
magnetic field squares in the region between the plates are simplified to%
\begin{align}
\left\langle E^{2}\right\rangle _{m\rightarrow 0}& =\frac{\left( 1-D\right)
\Gamma \left( \frac{D+1}{2}\right) }{\left( 4\pi \right) ^{\frac{D-1}{2}%
}a^{D+1}}\left[ \zeta \left( D+1\right) -\sum_{n=-\infty }^{\infty }\frac{3/2%
}{\left\vert n-z/a\right\vert ^{D+1}}\right] ,  \notag \\
\left\langle B^{2}\right\rangle _{m\rightarrow 0}& =\frac{(1-D)\Gamma \left( 
\frac{D+1}{2}\right) }{\left( 4\pi \right) ^{\frac{D-1}{2}}a^{D+1}}\left[
\zeta \left( D+1\right) +\sum_{n=-\infty }^{\infty }\frac{D-3/2}{\left\vert
n-z/a\right\vert ^{D+1}}\right] .  \label{EB2m0c}
\end{align}%
For the VEVs of the components of the energy-momentum tensor we get%
\begin{align}
\left\langle T_{l}^{l}\right\rangle _{m\rightarrow 0}& =\frac{\left(
1-D\right) \Gamma \left( \frac{D+1}{2}\right) }{\left( 4\pi \right) ^{\frac{%
D+1}{2}}a^{D+1}}\left[ \zeta \left( D+1\right) +\sum_{n=-\infty }^{\infty }%
\frac{(D-3)/2}{\left\vert n-z/a\right\vert ^{D+1}}\right] ,  \notag \\
\left\langle T_{D}^{D}\right\rangle _{m\rightarrow 0}& =\frac{D\left(
D-1\right) \Gamma \left( \frac{D+1}{2}\right) }{\left( 4\pi \right) ^{\frac{%
D+1}{2}}a^{D+1}}\zeta (D+1).  \label{Tllm0c}
\end{align}%
The limiting VEVs (\ref{EB2m0c}) and (\ref{Tllm0c}) coincide with the
corresponding expectation values for a massless vector field. They are
obtained from the results in \cite{Saha20} for the AdS bulk in the limit of
the infinite curvature radius. In the special case $D=3$, from (\ref{Tllm0c}%
) we get the Casimir result $p_{\mathrm{C}}=-\pi ^{2}/(240a^{4})$ for the
vacuum pressure. In this case, the Casimir energy density is zero in the
regions $z<0$ and $z>a$ and its distribution in the region between the
plates is uniform. Figure \ref{figT00cm0} presents the dependence of the
Casimir energy density (in units of $1/a^{D+1}$) for a massless vector field
with PEC boundary conditions on the ratio $z/a$ in spatial dimensions $%
D=3,4,5,6$. The modulus of the combination $a^{D+1}\left\langle
T_{0}^{0}\right\rangle $ increases with increasing $D$. 

As we have seen, an important difference between the two types of boundary
conditions discussed above is that the PMC conditions constrain all the
polarization states of the Proca field, whereas the longitudinal
polarization modes are not influenced by PEC condition. As a consequence,
for the first case, the regions separated by the reflecting surfaces
(hyperplanes $z=0$ and $z=a$ in the present problem) are causally
independent and the local characteristics of the vacuum state in the given
region do not depend on the physical conditions in other regions. For
example, we can consider a problem involving a single, finite-thickness PMC
plate occupying the spatial region $0\leq z\leq L$. In this geometry, the
VEVs in the regions $z<0$ and $z>L$ (the vacuum regions) will be expressed
by the formulas given above for PMC conditions (with $a$ replaced by $L$).
The PEC interfaces are transparent to the longitudinal polarization modes of
the Proca field, so the VEVs in separate regions are sensitive to the
physical conditions in other regions. In this case, the longitudinal
polarization modes will contribute to the Casimir densities. Examples for
the total Casimir energy and forces in the problem of two parallel plates of
finite thickness have been studied in \cite{Bart85,Teo10,Teo12,Eder08}.

\section{Conclusion}

\label{sec:Conc}

We have studied the Casimir effect for a massive vector field in a\ general
number of spatial dimensions. The geometry of two parallel plates is
considered with two types of boundary conditions. The first one, referred to
as PMC condition, is a higher dimensional generalization of the boundary
conditions used in bag models of hadrons to confine the gluon fields inside
a finite volume and the second one (PEC condition) generalizes the boundary
condition imposed on the surface of perfect conductors in Maxwell
electrodynamics. The massive vector field in $(D+1)$-dimensional spacetime
has $D$ independent polarization states: $D-1$ states with transversal
polarization and one state of longitudinal polarization. The PMC conditions
constrain all the polarization states, whereas the PEC conditions do not act
on the longitudinal polarization. Due to this feature, the massless limit of
the vacuum local characteristics is significantly different for those
boundary conditions.

In the problem under consideration, all the properties of the vacuum state
are encoded in two-point functions. As such we have taken the positive
frequency Wightman functions for the vector potential and the field tensor.
The evaluation procedure is based on the direct summation of the
corresponding mode sums over a complete set of mode functions obeying the
boundary conditions. In the region between the plates, the mode functions
for the PMC conditions are given by (\ref{Atr}) and (\ref{Al}) with the
eigenvalues for the normal component of the wave vector given by (\ref{kD})
(excluding the mode $n=0$ for the longitudinal polarization). The two-point
function for the vector potential is expressed as (\ref{AA3}), where the
separate terms in the summation are defined by (\ref{ADD2}). It is
decomposed into three contributions (\ref{AAdec}), corresponding to the
boundary-free geometry, the part induced by a single boundary, and the
contribution generated by the second boundary. The massless limit of the
two-point function for the vector potential is singular: it diverges like $%
1/m^{2}$ (see (\ref{AA3m0})). The two-point function for the field tensor in
the problem with PMC conditions is given by (\ref{FF2p1}) and (\ref{FF2p})
and its massless limit is finite (see (\ref{FFm0lim})).

The Casimir densities are obtained from the two-point functions in the
coincidence limit of the spacetime arguments. For points outside the plates,
the divergences are contained in the parts corresponding to the
boundary-free problem. In our representations of the two-point functions
those parts are explicitly separated and the renormalization is reduced to
the subtraction of them. As important local characteristics of the vacuum
state, we have considered the expectation values of the electric and
magnetic field squares, of the energy-momentum tensor, and the vector field
condensate. The corresponding VEVs in the geometry of a single plate are
obtained in the limit $a\rightarrow \infty $. The single plate contributions
dominate in the total VEVs for points near the plates. For PMC boundary
conditions the VEV of the electric field squared is negative. The
corresponding Casimir-Polder forces, acting on a polarizable microparticle,
are repulsive with respect to the nearest plate. The VEV of the magnetic
field squared vanishes for $D=1$ and is positive in dimensions $D\geq 2$.
The vacuum energy density in the region between the plates is negative for $%
D=1,2$ and positive for $D\geq 3$. It is uniformly distributed in the
spatial dimension $D=2$. The vacuum energy density in the regions $z<0$ and $%
z>a$ is negative for $D=1$ and positive for $D\geq 3$. In those regions, the
vacuum energy-momentum tensor vanishes for $D=2$. The Casimir forces are
attractive for all separations between the plates. For a massive field, they
exponentially decay at large separations.\ We have also provided the
expressions for the VEVs in the massless limit. The VEV of the
energy-momentum tensor in that limit is traceless for $D=1,2$ and has a
nonzero trace for $D\geq 3$.

To compare the VEVs in the zero-mass limit of the Proca field, obeying PMC
boundary conditions, with the corresponding expectation values for a
massless vector field, we have separately studied the two-point functions
and the Casimir densities in the massless case. The VEVs of the electric and
magnetic field squares, as well as the condensate, are entirely determined
by the two-point function of the field tensor. In the zero-mass limit, these
VEVs coincide with those for a massless field. This is not the case for the
VEV of the energy-momentum tensor which also receives the contribution from
the two-point function of the vector potential. The longitudinal
polarization is absent for a massless field, whereas its contribution to the
vacuum energy-momentum tensor for the Proca field does not vanish in the
zero-mass limit.

We have also considered the two-point functions and the Casimir densities
for the Proca field with the PEC boundary conditions on the plates. These
conditions do not constrain the mode with longitudinal polarization and the
latter does not contribute to the Casimir parts in the VEVs. The transverse
polarization modes are given by (\ref{Atrc}) with the quantized normal
component of the wave vector (\ref{kDc}). For the PEC conditions the
two-point function of the vector potential is finite in the zero-mass limit.
Consequently, the Casimir densities in the zero mass limit of the Proca
field coincide with the corresponding VEVs for a massless vector field. The
vacuum energy density is negative for $D\geq 3$, positive for $D=2$, and
vanishes for $D=1$. The Casimir forces are attractive in spatial dimensions $%
D\geq 2$ and become zero for $D=1$. For $D\geq 2$ the ratio of the Casimir
forces for PMC and PEC boundary conditions is equal to the ratio of the
number of independent polarizations influenced by those conditions. The VEV
of the electric field squared for PEC conditions is positive for $D\geq 2$
and the corresponding Casimir-Polder forces are attractive with respect to
the nearest plate.

\section*{Acknowledgments}

The work was supported by the grants No. 21AG-1C047 and No. 24AA-1C025 of
the Higher Education and Science Committee of the Ministry of Education,
Science, Culture and Sport RA.

\appendix

\section{Transformation of the function $A_{j}(x,x^{\prime })$}

\label{sec:App1}

In this Appendix we present the transformation of the two-point function (%
\ref{Ajxx}). The summation over $n$ in (\ref{Ajxx}) is done by using the
Abel-Plana formula%
\begin{equation}
\sideset{}{'}{\sum}_{n=0}^{\infty }f(n)=\int_{0}^{\infty
}dy\,f(y)+i\int_{0}^{\infty }dy\,\frac{f(iy)-f(-iy)}{e^{2\pi y}-1},
\label{APF}
\end{equation}%
with the function%
\begin{equation}
f(y)=\frac{e^{-i\omega t}}{\omega }\cos \left( b_{j}y\right) ,\;\omega =%
\sqrt{\omega _{\mathbf{k}}^{2}+(\pi y/a)^{2}},\;\omega _{\mathbf{k}}=\sqrt{%
\mathbf{k}^{2}+m^{2}}.  \label{fy}
\end{equation}%
For the series in (\ref{Ajxx}) one has $b_{j}=|z+jz^{\prime }|$. Assuming
that $b_{j}+|\Delta t|<2a$, one gets 
\begin{equation}
\sideset{}{'}{\sum}_{n=0}^{\infty }\frac{e^{-i\omega \Delta t}}{\omega }\cos
\left( b_{j}k_{D}\right) =\frac{a}{\pi }\int_{0}^{\infty }du\,\cos \left(
b_{j}u\right) \frac{e^{-i\sqrt{u^{2}+\omega _{\mathbf{k}}^{2}}\Delta t}}{%
\sqrt{u^{2}+\omega _{\mathbf{k}}^{2}}}+\frac{2a}{\pi }\int_{\omega _{\mathbf{%
k}}}^{\infty }du\,\frac{\cosh \left( \sqrt{u^{2}-\omega _{\mathbf{k}}^{2}}%
\Delta t\right) }{\sqrt{u^{2}-\omega _{\mathbf{k}}^{2}}\left(
e^{2au}-1\right) }\cosh \left( b_{j}u\right) .  \label{Sn1}
\end{equation}%
In the first integral we write $\cos \left( b_{j}u\right)
=(e^{ib_{j}}+e^{-ib_{j}})/2$ and rotate the integration contour in the
complex $u$-plane by the angle $\pi /2$ for the part with $e^{ib_{j}}$ and
by the angle $-\pi /2$ for the part with $e^{-ib_{j}}$. In this way it can
be seen that%
\begin{equation}
\int_{0}^{\infty }du\,\cos \left( b_{j}u\right) \frac{e^{-i\sqrt{%
u^{2}+\omega _{\mathbf{k}}^{2}}\Delta t}}{\sqrt{u^{2}+\omega _{\mathbf{k}%
}^{2}}}=\int_{\omega _{\mathbf{k}}}^{\infty }du\frac{\cosh \left( \sqrt{%
u^{2}-\omega _{\mathbf{k}}^{2}}\Delta t\right) }{\sqrt{u^{2}-\omega _{%
\mathbf{k}}^{2}}}e^{-b_{j}u}.  \label{Reli}
\end{equation}%
By using this relation in (\ref{Sn1}) we obtain%
\begin{equation}
\sideset{}{'}{\sum}_{n=0}^{\infty }f(n)=\frac{a}{\pi }\int_{\omega _{\mathbf{%
k}}}^{\infty }du\frac{\cosh (\sqrt{u^{2}-\omega _{\mathbf{k}}^{2}}\Delta t)}{%
\sqrt{u^{2}-\omega _{\mathbf{k}}^{2}}}\left[ e^{-b_{j}u}+\frac{2\cosh \left(
b_{j}u\right) }{e^{2au}-1}\right] .  \label{Sn2}
\end{equation}%
This gives%
\begin{equation}
A_{j}(x,x^{\prime })=\frac{1}{\pi }\int \frac{d\mathbf{k\,}e^{i\mathbf{k}%
\Delta \mathbf{x}_{\parallel }}}{\left( 2\pi \right) ^{D-2}}\int_{\omega _{%
\mathbf{k}}}^{\infty }du\frac{\cosh (\sqrt{u^{2}-\omega _{\mathbf{k}}^{2}}%
\Delta t)}{\sqrt{u^{2}-\omega _{\mathbf{k}}^{2}}}\left[ e^{-b_{j}u}+\frac{%
2\cosh \left( b_{j}u\right) }{e^{2au}-1}\right] .  \label{Ajxx2}
\end{equation}

For the integration over the angular part of the vector $\mathbf{k}$, we
introduce in (\ref{Ajxx2}) spherical coordinates in the momentum space with
the polar axis along the vector $\Delta \mathbf{x}_{\parallel }$. The
integral over the angles is expressed in terms of the Bessel function $%
J_{(D-3)/2}(k|\Delta \mathbf{x}_{\parallel }|)$, where $k=|\mathbf{k}|$. As
the next step we introduce a new integration variable $w=\sqrt{u^{2}-\omega
_{\mathbf{k}}^{2}}$ and then pass to the polar coordinates in the plane $%
(w,k)$. The integral over the corresponding polar angle is evaluated by
using the formula \cite{Prud12} 
\begin{equation}
\int_{0}^{1}dx\,x^{\nu +1}\frac{\cosh \left( b\sqrt{1-x^{2}}\right) }{\sqrt{%
1-x^{2}}}J_{\nu }\left( cx\right) =\sqrt{\frac{\pi }{2}}c^{\nu }\frac{J_{\nu
+1/2}(\sqrt{c^{2}-b^{2}})}{\left( c^{2}-b^{2}\right) ^{\frac{2\nu +1}{4}}}.
\label{Int2}
\end{equation}%
As a result, the representation 
\begin{equation}
A_{j}(x,x^{\prime })=\frac{\left( 2\pi \right) ^{1-\frac{D}{2}}}{(-\Delta
x_{l}\Delta x^{l})^{\frac{D-2}{4}}}\int_{0}^{\infty }dr\,r^{\frac{D}{2}}J_{%
\frac{D}{2}-1}(r\sqrt{-\Delta x_{l}\Delta x^{l}})\frac{1}{u}\left[
e^{-b_{j}u}+\frac{2\cosh \left( b_{j}u\right) }{e^{2au}-1}\right] _{u=\sqrt{%
r^{2}+m^{2}}}  \label{Ajxx3}
\end{equation}%
is obtained, where $\Delta x_{l}\Delta x^{l}=\left( \Delta t\right)
^{2}-|\Delta \mathbf{x}_{\parallel }|^{2}$. An alternative representation is
obtained by using the relation%
\begin{equation}
\frac{2\cosh \left( b_{j}u\right) }{e^{2au}-1}=\sum_{s=\pm
1}\sum_{n=1}^{\infty }e^{-\left( 2na-sb_{j}\right) u}.  \label{exp2}
\end{equation}%
With this expansion, the integral over $r$ is evaluated by the formula \cite%
{Prud12} 
\begin{equation}
\int_{0}^{\infty }dr\,r^{D/2}J_{\frac{D}{2}-1}(r\sqrt{-\Delta x_{l}\Delta
x^{l}})\frac{e^{-c\sqrt{r^{2}+m^{2}}}}{\sqrt{r^{2}+m^{2}}}=\sqrt{\frac{2}{%
\pi }}m^{D-1}(-\Delta x_{l}\Delta x^{l})^{\frac{D-2}{4}}f_{\frac{D-1}{2}}(m%
\sqrt{c^{2}-\Delta x_{l}\Delta x^{l}}),  \label{Int3}
\end{equation}%
with the function (\ref{fnu}). As a result, the representation 
\begin{equation}
A_{j}(x,x^{\prime })=\frac{2m^{D-1}}{\left( 2\pi \right) ^{\frac{D-1}{2}}}%
\sum_{n=-\infty }^{+\infty }f_{\frac{D-1}{2}}(m\sqrt{\left( 2na-z-jz^{\prime
}\right) ^{2}-\Delta x_{l}\Delta x^{l}}),  \label{Ajxx4}
\end{equation}%
is obtained.

\end{document}